\documentclass[aps,prl,paper,reprint,amsmath,amssymb,superscriptaddress,floatfix,showpacs,showkeys,subeqn,footinbib]{revtex4-1} 
\usepackage{amsmath, amsfonts, graphics, graphicx, hyperref,url,array, multirow, subfigure, rotating,booktabs}
\usepackage{todonotes}
\usepackage{color}

\usepackage{longtable,capt-of}

\newcommand{\blue}[1]{\textcolor{blue}{#1}}
\newcommand{\beq}{\begin{equation}}
\newcommand{\eeq}{\end{equation}}
\newcommand{\kbar}{\bar{k}}
\newcommand{\kb}{\bar k}

\begin{document}

\setlength{\heavyrulewidth}{0.08em}
\setlength{\lightrulewidth}{0.05em}
\setlength{\cmidrulewidth}{0.03em}
\setlength{\aboverulesep}{0.4ex}
\setlength{\belowrulesep}{0.65ex}

\title{The maximum packing fraction of disordered bidisperse discs, \\
theory and predictions}
\author{Raphael Blumenfeld}
\affiliation{Gonville \& Caius College, University of Cambridge, Trinity St., Cambridge CB2 1TA, UK}

\date{25/01/2026}

\begin{abstract}
Predicting the highest packing fraction that disordered packing of bidisperse discs can achieve, $\phi_{max}$, is an important unresolved problem. 
Such a predictive theory is hindered by the density dependence on the packing protocol and by ensuring disorder. In bidisperse packings, this is more complicated because $\phi_{max}$ is not a number but a function of the disc species size ratio, $D$, and one of the disc species concentration, $p$. 
These difficulties have been overcome in monodisperse disc packings by using the cell order distribution (COD)~\cite{Bl21}.
Focusing on size ratios that prohibit rattlers, a COD-based theory is constructed here. It introduces bidisperse statistical closure equations, which combine an explicit disorder criterion with consistent cell and edge statistics, and yields $\phi_{max}(p,D)$ and structural predictions.
The a-priori disorder-guaranteeing criterion proposed here ensures fast exponential a decay of the occurrence probability of ordered clusters with size. Using a constrained maximum-entropy analysis to ensure maximum disorder, the occurrence probabilities of all the possible cell and edge configurations are derived in terms of $p$ and $D$ and they are found to be binomial.  All the results are assembled to yield an exact prediction of $\phi_{max}(p,D)$. 

\end{abstract}

\maketitle


\noindent\underline{\textbf{Introduction}}

Packing objects densely is a goal of practical importance in technology and more generally in human society. The quest to predict the packing fraction of the densest possible packing of a given number of objects focused much research. Special attention has been given to a canonical version of the problem - predicting the highest packing fraction of disordered packings of spheres in three dimensions  ($d=3$) and discs in two ($d=2$), named the random close packing (RCP) problem. This problem is relevant to many technological applications~\cite{Toetal10}, to liquid-solid phase transitions~\cite{Bernal}, to information packing~\cite{IT1,IT2,IT3}, to process and nuclear engineering~\cite{Proc,Nucl}, as well as to the understanding and modelling of many natural phenomena.  
A successful theoretical approach to solving the canonical problem has the potential to pave the way to predicting and facilitating packing of more general objects. 
Many experimental and numerical studies put the highest packing fractions of same-size (monodisperse) spheres and discs, $\phi_{RCP}$, inside narrow ranges of values, e.g., $0.82-0.86$ for monodisoperse discs in $d=2$~\cite{Kaetal71,QuTa74,BiTr84,Hietal90,Oheetal01}. All those studies use packings generated physically or numerically, forming a collection of trial-and-error cases. However, large as this collection is, it does not rule out a yet-untried process that could generate a denser state. 
However, there appears to be a controversy on the very definition of this problem.  For some, $\phi_{RCP}$ is the highest packing fraction that can be achieved for a given set of objects by any packing protocol, while for others this is the highest possible packing fraction of any disordered packing of such objects, irrespective of whether it can be generated by any protocol. The latter interpretation makes it an upper bound on any disordered packing packing fraction. It is the latter interpretation that is studied here and to distinguish it from the former, this packing fraction is called $\phi_{max}$.

Two main obstacles hinder a theoretical prediction. One is that, given any set of objects, there are many ways to pack them, each giving rise to a different packing fraction. As each generating process has parameters that can be varied continuously, the problem exists in an infinite-dimensional parameter space~\cite{Bl21}, which no trial-and-error approach can exhaustively explore. 
Another obstacle, especially prominent in sphere and disc problems, is ensuring disorder - under most packing processes, these objects tend to 'crystalise' into ordered structures that are denser than disordered ones - hexagonal in $d=3$ and trigonal in $d=2$. Thus, deriving $\phi_{max}$ theoretically requires an  {\it a-priori} criterion that ensures disorder. While several disorder criteria have been proposed in the literature, e.g., in~\cite{ZePr27,Stetal83,Toetal00,Jietal11,Moetal24}, those are often a-posteriori in the sense that they require testing disorder in already-existing packings. This means that a packing has to be generated first, which clashes with the infinite-parameter-space difficulty. 
It was shown that both these obstacles can be circumvented in $d=2$ packings of monodisperse discs by using the cell order distribution (COD)~\cite{Bl21}. 

While the prediction of $\phi_{max}$ appears reasonably resolved for monodisperse discs in $d=2$~\cite{Bl21}, to minimise crystallisation, many simulations and experiments are carried out on packings of bidisperse discs, e.g.,~\cite{ViBo72,Bietal86,OkOd04,Xuetal05,Koetal16,Br25,Meetal24}. Therefore, a prediction of $\phi_{max}$ in such systems is, arguably, more useful. 
This problem is more complex than the monodisperse one because, in addition to the aforementioned difficulties, $\phi_{max}$ in this case is a function of the disc sizes ratio, $D$, and the concentration of the smaller discs, $p$. 

Previous theoretical predictions of disordered disc-packing densities use small $D$-differences, statistical geometry, liquid-state and replica approximations, or mappings between monodisperse and mixture packing fractions~\cite{Ber83,Bietal86,PaZa10,XuRi11,Tietal15,SaLo26,Br24,Br26}.
The present approach connects the density directly to the topology and geometry of the contact network, combining an explicit constraint on ordered-cluster occurrence probabilities (OPs) with two constrained maximum-entropy calculations for cell and edge configurations to ensure maximum disorder. Within the stated assumptions, it determines $\phi_{max}(p,D)$ together with edge-type probabilities and the mean coordination numbers of each disc species. The latter result avoids existing models that require an empirical monodisperse packing fraction or impose isostatic conditions. The advantage of the current theory is in linking the maximum density, a disorder criterion, and independently measurable contact-network statistics.

It is sufficient to consider the size ratio of the two disc species, $1:D$, as scaling both sizes by the same factor does not affect the packing fraction. 
In the following, this ratio is limited to $1<D<D_{max}=\sqrt{3}/(2-\sqrt{3})$, to ensure no `rattlers' can nestle inside enclosures of either three or four discs in contact. 
The packings are assumed isotropic and to include a very large number of discs, $N\to\infty$. These assumptions ensure that all possible structural packing and cell configurations are realised and that boundary corrections can be neglected. 

The paper is organised as follows. The COD is defined in the next section and its usefulness to solving this problem is described. 
By showing that the packing fraction increases with decreasing mean cell order, it is argued that $\phi_{max}$ must emerge in bidisperse packings in which the highest cell order is four. 
An {\it a-priori} disorder criterion is then proposed, which limits the OP of equilateral cells. It also makes it possible to derive the range $p(D)$, within which bidisperse packings are certain to be disordered. To derive $\phi_{max}$, calculations of cell configurations and disc area statistics are required and these are done next. A key calculation is of the required OPs of each configuration of cells of orders three and four (henceforth $3$- and $4$-cells) and a closed set of equations and constraints is constructed to this end. These yield: the highest admissible fraction of $3$-cells to avoid order, the OPs of the different edge types in the contact network graph, and the specific mean coordination numbers of the small and large discs. Finally, using the cell configuration OPs, $\phi_{max}$ is derived as a function of $p$ and $D$. The paper is concluded with a summary of the results and a discussion of potential uses and extensions. \\

\noindent\underline{\textbf{The COD}}

The following analysis focuses on isotropic packings of $N\gg1$ discs, of which a fraction $p$ are unit discs and $1-p$ of diameter $D$.
In any jammed disc packing, connecting the centres of each disc with those it is in contacts with by 'edges' generates a graph, whose nodes are the disc centers. The smallest voids enclosed by the edges are the 'cells', also known as the elementary loops. The number of edges surrounding a cell is defined as its order. Given $N_k$ cells of order $k$, constituting a fraction $Q_k=N_k/N_c$ of all the graph's cells, $N_c\gg1$, the distribution of the these fractions is the COD~\cite{MaBl14,Waetal20}. 

One advantage of the COD is that it can be used to classify all possible packing, irrespective of the packing protocol, as any packing process generates one. Another advantage is that the packing fraction can be calculated directly from it, as will be shown below. These make the COD very useful for addressing this problem - it translates to finding the COD that provides the highest packing fraction, $\phi_{max}$ . This route circumvents the difficulty of the infinite process parameter space~\cite{Bl21} because the solution to this problem is independent of the generating protocol. It should be clarified that different protocols for packing isotropic systems of $N\gg1$ discs with the same $p$ and $D$, could, in principle, give rise to the same COD, in spite of the packings having different mechanical properties. Nevertheless, if all their structural configurations are realised then they would all have {\it the same packing fraction}. 

That the problem translates to that of finding the 'ideal' COD can be seen as follows. Given a given collection of $N$ rigid discs, irrespective of either their other material properties (e.g., inter-particle friction) or the protocol used to pack them, the packing fraction is always a monotonically increasing function of  the mean number of contacts per disc, $\bar{z}$~\cite{Uretal06,MaBl14,Be15}, $d\phi/d\bar{z}>0$. In turn, $\bar{z}$ is related to the mean cell order, 
\beq
\kbar = \sum_{k=3}^K k Q_k \ ,
\label{kbar} 
\eeq
with $K$ the highest cell order in the packing. To obtain this relation, start from Euler's topological relation for the aforementioned contact planar graphs, 
\beq
N-N_e+N_c=1 \ ,
\label{Euler1}
\eeq
where $N_e$ the number of edges. As each edge connects two disc centres, $N_e=\bar{z} N/2$, and together with (\ref{Euler1}) yields the ratio
\beq
\frac{N_c}{N} = \frac{\bar{z}}{2} - 1 \ ,
\label{Euler2}
\eeq
up to a negligible correction of $1/N$. But each edge is also shared by two cells, $N_e=\kbar N_c/2$ and combining this with (\ref{Euler1}) yields  
\beq
\frac{N}{N_c} = \frac{\kbar}{2} - 1 \ .
\label{Euler3}
\eeq
Thus, up to negligible boundary corrections, 
\begin{equation}
\frac{\bar{k}-2}{2}= \frac{2}{\bar{z} - 2} 
\label{kz0}
\end{equation}
and $\kbar$ decreases monotonically with $\bar{z}$. 
It follows that 
\beq
\frac{\partial\phi}{\partial\kbar} = \frac{\partial\phi}{\partial\bar{z}}\frac{\partial\bar{z}}{\partial\kbar} < 0
\label{kz}
\eeq
and the packing fraction decreases monotonically with $\kbar$. This observation bears directly on the problem at hand - it means that the highest possible packing fraction corresponds to the COD with the lowest mean cell order. While this means that the highest packing fraction would be achieved for packings that contain only $3$-cells, not all such packings are disordered - they may include large ordered trigonal regions, especially when $p\to0$ or $p\to1$. 
While lowering $\kbar$ increases the packing fraction, it also increases the probability of ordered clusters. The 'ideal' COD is the optimal compromise between these two competing effects and finding it is the objective here. As will be shown later, it need not include cells of orders higher than $4$ and, consequently, the following analysis focuses on CODs that have only $3$- and $4$-cells. \\

\noindent\underline{\textbf{The configurational statistics}}

\begin{figure}[h]
\includegraphics[width=.45\textwidth]{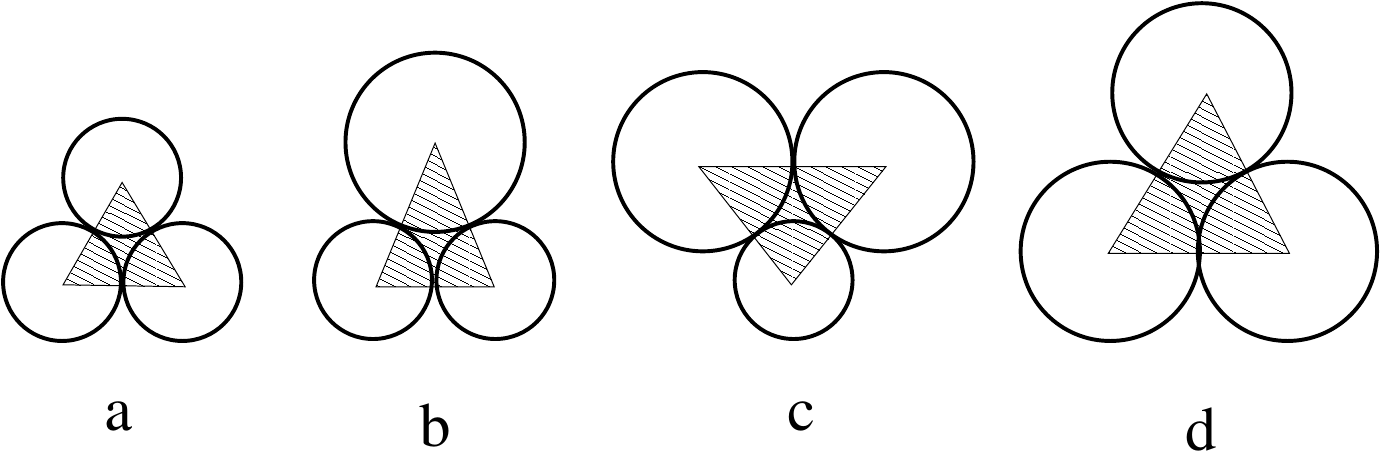}
\caption{The four possible $3$-cell configurations, $i=a,b,c,d$, and their areas, $S_{3i}$ (shaded). $S^{disc}_{3i}$ are the areas that the discs occupy within the shaded areas.}
\label{fig:3combs}
\end{figure}

\begin{figure}[h]
\includegraphics[width=.9\linewidth]{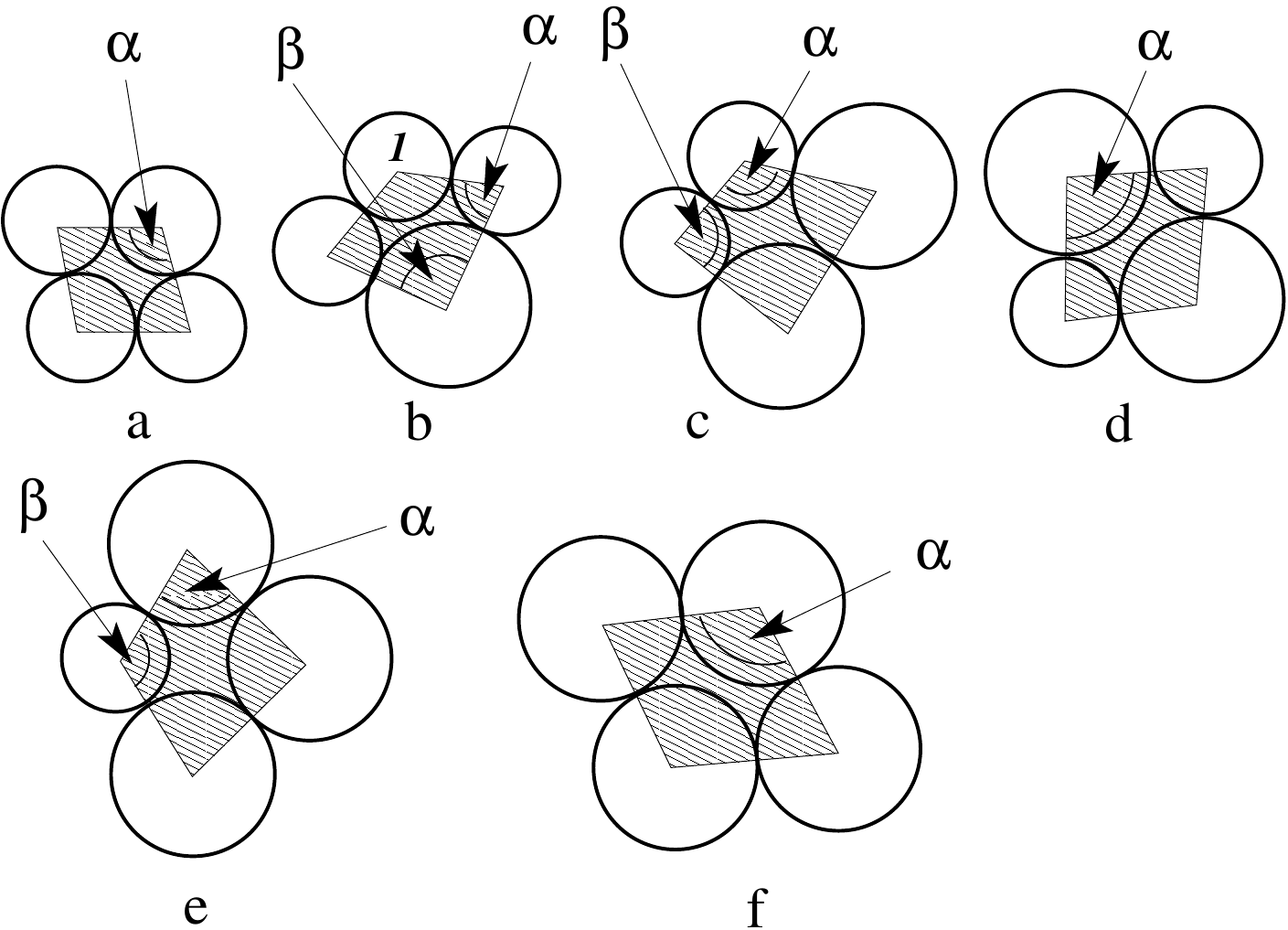}
\caption{The six possible 4-cell configurations, $S_{4i}$, $i=a,b,c,d,e,f$, and their areas (shaded). $S^{disc}_{4i}$ are the areas the discs occupy within the shaded cell areas..}
\label{fig:4combs}
\end{figure}

All the possible configurations of $3$- and $4$-cells are shown in FIgs.~\ref{fig:3combs} and \ref{fig:4combs}. Each $k$-cell configuration consists of a unique sequence of $m$ small discs and $k-m$ large ones around it. The assumptions of isotropy and $N_c\to\infty$ ensure that all these configurations are realised in the packing and that the occurrence probability of each configuration, $p_{ki}$ ($k=3,4$) is fixed given $p$ and $D$. 
Given a sequence of discs around a $k$-cell, its geometry depends of $k-3$ internal angles. Thus, the area of an $i$th $3$-cell configuration, $S_{3i}$, is fixed, while the area of an $i$th $4$-cell configuration, $S_{4i}$, depends on one internal angle, e.g., either $\alpha$ or $\beta$, in Fig.~\ref{fig:4combs}. The mean area of the $i$th $4$-cell configuration is then 
\beq
\langle S_{4i}\rangle = \frac{1}{\alpha_{i,max}-\alpha_{i,min}}\int\limits_{\alpha_{i,min}}^{\alpha_{i,max}} S_{4i} P(\alpha) d\alpha \ .
\label{int4}
\eeq
The angles $\alpha$ and $\beta$ can vary in the ranges shown in Table~\ref{table1},
\begin{table}[h]
\caption{The ranges of the $4$-cell configurations' internal angles} 
\centering 
\begin{tabular}{c c c c c} 
\hline\hline 
Case & $\alpha_{min}$ & $\alpha_{max}$ & $\beta_{min}$ & $\beta_{max}$ \\ [0.5ex] 
\hline 
4a & $\frac{\pi}{3}$ & $\frac{2\pi}{3}$ & irrelevant & irrelevant  \\ 
4b & AC & AC $+\frac{\pi}{3}$ & $2$AS & $4$AS  \\
4c & AC & AC $+2$BS  & AC & AC $+2$BS  \\
4d & $2$AS & $2$BC & irrelevant & irrelevant  \\
4e & BC & BC $+ \frac{\pi}{3}$ & $2$BS & $4$BS  \\
4f & $\frac{\pi}{3}$ & $\frac{2\pi}{3}$  & irrelevant & irrelevant \\ [1ex] 
\hline 
\end{tabular}
\label{table1} 
\end{table}
where, for brevity, the following have been defined:
\begin{align}
AS&\equiv\arcsin\frac{1}{D+1} \quad ;\quad AC\equiv\arccos\frac{1}{D+1} \notag \\
BS&\equiv\arcsin\frac{D}{D+1} \quad ;\quad BC\equiv\arccos\frac{D}{D+1} \ .
\end{align}
The total areaa per cell, which each of these subsets of cells occupy, are 
\beq
\bar{S}_{3} = Q_3\sum_{i=a}^d p_{3i} S_{3i} \qquad , \qquad \bar{S}_{4} = Q_4\sum_{i=a}^f p_{4i} \langle S_{4i}\rangle \ .
\label{MeanAreas34}
\eeq
With the general total packing area of any packing being 
\begin{align}
S_{pack} = N_c\sum_{k=3}^\infty Q_k\bar{S}_{k}  \ ,
\label{Spack}
\end{align} 
that of the $3/4$  packings is $S_{pack} = N_c\left(\bar{S}_{3}+\bar{S}_{4}\right)$.
The cell configurations areas are straightforward to calculate from geometric considerations:
\begin{align}
S_{3a} & = \frac{\sqrt{3}}{4} \quad ; \quad S_{3b} = \frac{\sqrt{D\left(D+2\right)}}{4} \notag \\
S_{3c} & =  \frac{D\sqrt{1+2D}}{4} \quad ; \quad S_{3d} = \frac{\sqrt{3}}{4}D^2 \ ,
\label{eq:3areas}
\end{align}
and
\begin{align}
S_{4a} &= \sin{\alpha}  \quad ; \quad S_{4b} =  \frac{D+1}{2}\sin{\alpha} \notag \\
S_{4c} &=  \frac{(D+1)\sin{\frac{\alpha}{2}}}{4 BS} 
\left[\cos{\frac{\alpha}{2}} + \sqrt{D-\sin^2{\frac{\alpha}{2}}}\right]  \notag \\
S_{4d} &= \left(\frac{D+1}{2}\right)^2\sin\alpha   \quad ; \quad S_{4e} = \frac{D(D+1)}{2}\sin{\alpha}  \notag \\ 
S_{4f}  &= D^2\sin{\alpha} \ .
\label{eq:4CellAreas}
\end{align}
There is a direct relation between configurational entropy and the packing fraction; given a sequence of discs around  $k$-cell, the more elongated its shape the more it contributes to $\phi$ due to its lower area-to-perimeter ratio. The elongated shapes dominate the total number of possible cell configurations and, thence, the entropy. Indeed, it has been shown that, in the absence of mechanical stability constraints, the higher the entropy the higher the packing fraction~\cite{MaBl17,Suetal20}. Following this rationale, it is expected that $\phi_{max}$ should correspond to high-entropy distributions. Moreover, the higher the entropy the more disordered the packing. In particular, to calculate the averages of the areas in (\ref{eq:4CellAreas}), (\ref{int4}) is used with $P(\alpha)$ taken to be uniform, $P(\alpha)=1/\left(\alpha_{max}-\alpha_{min}\right)$, because it maximises the entropy in the absence of other constraints or information. \\
Integrating expressions (\ref{eq:4CellAreas}), the mean areas of the $4$-cell configurations are
\begin{align}
\langle S_{4a} \rangle & = \frac{3}{\pi} \notag \\
\langle S_{4b} \rangle & = \frac{3}{4\pi}\left[1 + \sqrt{3D(D+2)}\right] \notag \\
\langle S_{4c} \rangle & = \frac{D+1}{8\arcsin{\frac{D}{D+1}}}\Bigg[
\frac{1}{D+1}-\cos{\alpha_{max}} \notag \\
&+ 2(D-1)\ln{\frac{\cos{\frac{\alpha_{min}}{2}}+\sqrt{D-\sin^2{\frac{\alpha_{min}}{2}}}}
{\cos{\frac{\alpha_{max}}{2}}+\sqrt{D-\sin^2{\frac{\alpha_{max}}{2}}}}} \notag \\
&+ 2\cos{\frac{\alpha_{min}}{2}}\sqrt{D-\sin^2{\frac{\alpha_{min}}{2}}} \notag \\
&-2\cos{\frac{\alpha_{max}}{2}}\sqrt{D-\sin^2{\frac{\alpha_{max}}{2}}} \Bigg] \notag \\
\langle S_{4d} \rangle & = \frac{D/2}{\arccos{\left(\frac{D}{D+1}\right)}-\arcsin{\left(\frac{1}{D+1}\right)}} \notag \\
\langle S_{4e} \rangle & = \frac{3D}{4\pi}\left[D+\sqrt{3(2D+1)}\right]  \notag \\
\langle S_{4f} \rangle & = \frac{3D^2}{\pi}  \ .
\label{eq:S4CellareasAv}
\end{align}
Substituting (\ref{eq:3areas}) and (\ref{eq:S4CellareasAv}) into (\ref{MeanAreas34}) and (\ref{Spack}) yields $S_{pack}$. 

The total disc area is $S_{disc}=\left[p+(1-p)D^2\right]N\pi/4$, but there is another way to calculate it. A part of each cell configuration, $S_{ki}^{disc}$, is occupied by sections of the discs that surround it, with each disc section determined by the internal angle formed by the cell's edges at the disc centre, e.g., angles $\alpha$ and $\beta$ in the $4$-cell configurations shown in Fig.~\ref{fig:4combs}. As the internal angles of a $k$-cell sum up to $(k-2)\pi$, averaging over the disc sizes in all the $k$-cell configurations, the total area occupied by the discs within the $k$-cells is $N_c(k-2)\pi\left[p+(1-p)D^2\right]/8$.  Averaging over the COD, the packing fraction is then,
\begin{align}
\phi &= \frac{\sum_{k=3}^\infty Q_k \frac{(k-2)\pi}{8}\left[p+(1-p)D^2\right]}{S_{pack}}N_c \notag \\
&= \frac{\pi\left[p+(1-p)D^2\right]\left(\bar{k}-2\right)}{8S_{pack}} N_c \ ,
\label{Genphi}
\end{align}
where $\bar{k}=\sum_{k=3}^\infty kQ_k$ is the mean cell order. 
This expression is valid for any isotropic packing realisation, when $N_c\gg1$, establishing that all packings with the same $p$, $D$, and COD have the same packing fraction. \\
Relation (\ref{Genphi}) also shows that, given $p$ and $1<D<D_{max}$, the packing fraction increases as $\bar{k}$ decreases, in line with the argument presented above. This reduces the problem to that of finding the COD with the highest possible fraction of low-order cells, while keeping the packing disordered. \\

\noindent\underline{\textbf{Ensuring disorder}}

While several disorder measures and criteria have been suggested in the literature~\cite{ZePr27,Stetal83,Toetal00,Jietal11,Moetal24}, they are a-posteriori in the sense that they require generating a packing in order to measure its state of disorder. Measuring assemblies already lacked by a specific protocol misses the present aim at generality. It is therefore necessary to establish a criterion that ensures disorder independently of any protocol. Such an a-priori  criterion is proposed next. 

Order in bidisperse packings corresponds mainly to large clusters of $3$-cells of identical discs, configuration $3a$ or $3d$. The higher the total fraction of $3$-cells, $Q_3$, the higher the probability of such clusters occurring and they are inescapable when $p\to0$ or $1$. This competes with the observation that the higher $Q_3$ is the denser the packing. Thus, the objective is to identify the highest value of $Q_3$, $Q_{3,max}$, for which disorder is still assured and, from it, derive $\phi_{max}$ and the value of $p$, at which it occursl. 
The rationale underlying the proposed disorder criterion is to ensure that the OPs of ordered clusters decay exponentially fast with size, which is done here by limiting the OP of neighbouring identical equilateral cells to below $w=1/3$. It is shown below that, for the wide range of values of $p$ often used in experiments and simulation, this choice affects neither the ordered clusters OPs nor $\phi_{max}$.

It should be commented in passing that ordered structures of repeated lattice units can also occur for specific values of $p$ and $D$~\cite{Fl60,Fe25}. For example, one can construct arbitrarily large clusters consisting only of configuration $3b$ as a unit lattice. In this case, $p=2/3$ and $\phi=\left(AC+D^2AS\right)/\sqrt{D(D+2)}$~\cite{Fl60}. Such singular cases are disregarded here because the probability of $L$-size such clusters occurring spontaneously is diminishingly small, $<\left(Q_3p_{3b}\right)^L$, for $L>6$. 

Given the $3$-cells fraction, $Q_3$ and the OPs $p_{3a}$ and $p_{3d}$, the probability that a cell of either configuration has more than one identical neighbour is 
\begin{align}
R_{a} = &Q_3p_{3a}\left[ \left(Q_3p_{3a}\right)^3 + 3\left(1 - Q_3p_{3a}\right)\left(Q_3p_{3a}\right)^2\right] \notag \\
R_{d} = &Q_3p_{3d}\left[\left(Q_3p_{3d}\right)^3 + 3\left(1 - Q_3p_{3d}\right)\left[Q_3p_{3d}\right)^2\right] \ .
\label{criteria}
\end{align}
Imposing that both $R_{a}\leq w$ and $R_{d}\leq w$, the solution to (\ref{criteria}) is 
\beq
Q_{3,max} = {\rm min}\left\{\frac{A}{p_{3a}},\frac{A}{p_{3d}},1\right\}   \ .
\label{Q3max}
\eeq
with $A=0.5622..$ for $w=1/3$. 
For further reference, while all the numerical values in the following can be obtained to any precision, they will be abbreviated to $10^{-4}$ precision, for brevity. To determine $Q_{3,max}$, a derivation of $p_{3a}$ and $p_{3d}$ in terms of $p$ and $D$ are required, which is done next.  \\

\noindent\underline{\textbf{Cell configurations and edge OPs} }

In isolation, the OP of configuration $i$ of a $k$-cell, consisting of $m$ small discs and $k-m$ large ones, $p_{ki}$, is proportional to $p^m(1-p)^{k-m}$. However, in the packing, adjacent cells share two discs and the $p_{ki}$s need to be calculated more carefully. To take the sharing constraint into consideration, let us introduce the three OPs $e_{jm}$ ($j,m=1,D$) of an edge connecting the centres of discs of size $j$ with their neighbours of size $m$. In $3/4$ packings there are 10 \blue{$p_{ki}$s} and three $e_{jm}$s, as $e_{jm}=e_{mj}$. This section consists of a detailed calculation of these 13 \blue{OPs and a remarkable result is derived - all these OPs can all be parameterised by a single variable, in which they are binomial. }
The calculation also requires introduction of \blue{the fractions of} mean coordination numbers of the two disc sizes, $\bar{z}_1$, and $\bar{z}_D$. In each configuration, $ki$, the number of edges connecting discs of sizes $j$ and $m$, denoted as $n_{jm}(ki)$, are listed in Table~\ref{table2}. The table also lists the number of ways to arrange configuration $ki$, $g_{ki}$, which ensure that $n_{11}+n_{1D}+n_{DD}=k$ for each $k$-cell configuration. 
\begin{table}[h]
\caption{The number of edges in each configuration, $n_{jm}$, and the number of each configuration arrangements, $g_{ki}$} 
 \centering
\begin{tabular}{lccccc}
\hline\hline
         & discs sequence (cyclic) & $n_{11}$ & $n_{1D}$ & $n_{DD}$ & $g_{ki}$ \\
\hline
$3a$ & $111$ & 3 & 0 & 0 & 1 \\
$3b$ & $D11$ & 1 & 2 & 0 & 3 \\
$3c$ & $DD1$ & 0 & 2 & 1 & 3 \\
$3d$ & $DDD$ & 0 & 0 & 3 & 1 \\
$4a$ & $1111$ & 4 & 0 & 0 & 1 \\
$4b$ & $111D$ & 2 & 2 & 0 & 4 \\
$4c$ & $11DD$ & 1 & 2 & 1 & 4 \\
$4d$ & $1DDD$ & 0 & 4 & 0 & 2 \\
$4e$ & $1DDD$ & 0 & 2 & 2 & 4 \\
$4f$  & $DDDD$ & 0 & 0 & 4 & 1 \\
\hline
\end{tabular}
 \label{table2}
\end{table}

\blue{As one aims for highest disorder, the distributions of the $p_{ki}$s and the $e_{jm}$s must maximise} their entropies, subject to the following constraints:
\begin{align}
&Q_3 + Q_4 = 1 \label{11eq1} \\
&\sum_{i=a}^d p_{3i} = 1 \label{11eq2} \\
&\sum_{i=a}^f p_{4i} = 1\label{11eq3} \\
&e_{11}+e_{1D}+e_{DD}=1 \label{11eq4} \\
&\kbar e_{11} = Q_{3}\left(3p_{3a}+p_{3b}\right) + Q_4\left(4p_{4a}+2p_{4b}+p_{4c}\right)  \label{11eq5} \\
&\kbar e_{DD} = Q_{3}\left(p_{3c}+3p_{3d}\right) + Q_4\left(p_{4c}+2p_{4e}+4p_{4f}\right) \label{11eq6} \\
&p\bar{z}_1 = \left(e_{11} + \frac{1}{2} e_{1D}\right) \bar{z} \ . \label{11eq7}  
\end{align}
It should be noted that relations (\ref{11eq4}) and (\ref{11eq7}) imply $(1-p)\bar{z}_D=\left(e_{DD} + e_{1D}/2\right)\bar{z}$, ensuring that $p\bar{z}_1+(1-p)\bar{z}_D=\bar{z}$. 

As $N_c\to\infty$, the likeliest distribution of the $p_{ki}$s, as well as the one that is most disordered, is that of maximum entropy: 
\begin{align}
S_c &= - \sum_{k=3}^4 \sum_{i} g_{ki} \frac{Q_k p_{ki}}{g_{ki}} \ln{\frac{Q_k p_{ki}}{g_{ki}}} notag \\
& = -\sum_{k=3}^4 Q_k\ln{Q_k} - \sum_{i} Q_k p_{ki} \ln{\frac{p_{ki}}{g_{ki}}} \ .
\label{Sc}
\end{align}
It needs to be maximised at fixed $Q_3$, $n_{11}$, and $n_{DD}$, subject to the constraints (\ref{11eq2}), (\ref{11eq3}), (\ref{11eq5}), and (\ref{11eq6}). These constraints are assigned, respectively, the Lagrange multipliers: $\mu_3$, $\mu_4$, $\lambda_{11}$, and $\lambda_{DD}$.  The corresponding Lagrangian to maximise is therefore
\begin{align}
{\cal{L}}_c & = -\sum_{k=3}^4 Q_k\ln{Q_k} - \sum_{i} Q_k p_{ki} \ln{\frac{p_{ki}}{g_{ki}}} \notag \\
& - \mu_3\left(\sum_{i=a}^d p_{3i} - 1\right) - \mu_4\left(\sum_{i=a}^f p_{4i} - 1\right) \notag \\
& - \lambda_{11} \left[\sum_{k,i} Q_k n_{11}(ki) p_{ki} - \kbar e_{11}\right] \notag \\
& - \lambda_{DD} \left[\sum_{k,i} Q_k n_{DD}(ki) p_{ki} - \kbar e_{DD}\right] \ .
\label{Lc}
\end{align}
Solving by differentiations with respect to the $p_{ki}$s yields
\begin{align}
p_{ki} = g_{ki} e^{-1 - \mu_k/Q_k - \lambda_{11} n_{11}(ki) - \lambda_{DD} n_{DD}(ki) }  
\label{Lc1}
\end{align}
and the normalised OPs are 
\begin{align}
&\left(p_{3a}.p_{3b},p_{3c},p_{3d}\right) = \notag \\
&\frac{\left(e^{-3\lambda_{11}},3e^{-\lambda_{11}},3e^{-\lambda_{DD}},e^{-3\lambda_{DD}}\right)}{Z_3} \ ,
\label{p3lambda}
\end{align}
\begin{align}
&\left(p_{4a}.p_{4b},p_{4c},p_{4d},p_{4e},p_{4f}\right) =  \notag \\
&\frac{
\left(e^{-4\lambda_{11}},4e^{-2\lambda_{11}},4e^{-\lambda_{11}-\lambda_{DD}},2,4e^{-2\lambda_{DD}},e^{-4\lambda_{DD}}\right)}{Z_4}  \ ,
\label{p4nu}
\end{align}
with normalisation constants 
\begin{align}
Z_3 =&e^{-3\lambda_{11}}+3e^{-\lambda_{11}}+3e^{-\lambda_{DD}}+e^{-3\lambda_{DD}} \notag \\
Z_4 =&e^{-4\lambda_{11}}+4e^{-2\lambda_{11}}+4e^{-\lambda_{11}-\lambda_{DD}}+ \notag \\ 
&2+4e^{-2\lambda_{DD}}+e^{-4\lambda_{DD}} \ . \notag
\end{align}
The prefacors $e^{-1-\mu_k/Q_3}$ are determined by conditions (\ref{11eq2}) and (\ref{11eq3}).
Eliminating the Lagrange multipliers from the resulting ten equations, yields
\begin{align}
& p_{3a} p_{3c}^3 = p_{3d} p_{3b}^3 \label{Lc2} \\
& p_{4b}^2 = 8p_{4a} p_{4d} \label{Lc3} \\
& p_{4e}^2 = 8p_{4f} p_{4d} \label{Lc4} \\
& p_{4c}^2 = p_{4b} p_{4e} \label{Lc5} \\
& \frac{3p_{3a}}{p_{3b}} = \frac{p_{4b}}{2p_{4d}}.\label{Lc6} \\
& \frac{3p_{3d}}{p_{3c}} = \frac{p_{4e}}{2p_{4d}} \ .
\label{Lc7}
\end{align}
The entropy of the type of a randomly chosen edge is
\begin{align}
S_e &= - \sum_{j,m=1,D} e_{jm}\ln{\frac{e_{jm}}{g_{jm}}}  \ ,
\label{Se}
\end{align}
subject to conditions (\ref{11eq4}) and (\ref{11eq7}), to which $\nu_1$ and $\nu_2$ are assigned as Langrange multipliers, respectively. 
The appropriate Lagrangian is then
\begin{align}
{\cal{L}}_e & = -\sum_{j,m=1,D} e_{jm}\ln{\frac{e_{jm}}{g_{jm}}} - \nu_1\left(\sum_{j,m=1,D} e_{dim} - 1\right) \notag \\
& - \nu_2\left(e_{11} + \frac{1}{2}e_{1D} - \frac{\bar{z}_1}{p\bar{z}}\right)  
\label{Le}
\end{align}
and minimising it with respect to the $e_{jm}$s yields  
\beq
e_{jm} = e^{-1-\nu_1}
\begin{cases}
e^{-\nu_2} & jm = 11 \\
2e^{-\nu_2/2} & jm = 1D \\
1 & jm = DD  \ ,
\end{cases}
\label{Le2}
\eeq
which also gives 
\beq
e_{1D}^2 - 4e_{11}e_{DD} = 0 \ .
\label{Le3}
\eeq
Relations (\ref{11eq4})-(\ref{11eq6}) can be used to express all the $e_{jm}$s in terms of $e^{-\lambda_{11}}$ and $e^{-\lambda_{DD}}$. Multiplying (\ref{Le3}) by $\left(\kbar Z_3Z_4\right)^2$ and substituting these expressions into it, the left hand side can be factorised into 
$$\left(1 - e^{-\lambda_{11}-\lambda_{DD}}\right)P\left(e^{-\lambda_{11}},e^{-\lambda_{DD}},Q_3,Q_4\right) \ ,$$
where $P$ is quadratic in $\left(Q_3,Q_4\right)$ and each of the coefficients multiplying $Q_3^2$, $Q_3Q_4$, and $Q_4^2$ are polynomials in the variables $e^{-\lambda_{11}}$ and $e^{-\lambda_{DD}}$. 
This can be readily checked by substituting $e^{-\lambda_{DD}}=e^{\lambda_{11}}$ into (\ref{Le3}), expressed in terms of the exponentials and observe that this makes it vanish. 
Additionally, the polynomial $P$ is derived in detail in the Supplemental material~\cite{SM} and it is shown to be positive definite, with no real roots. Thus, the only solution to eq. (\ref{Le3})  is that $\lambda_{DD}=-\lambda_{11}$.  
Plugging this into (\ref{p3lambda}) and (\ref{p4nu}) yields  
\begin{align}
&\left(p_{3a}.p_{3b},p_{3c},p_{3d}\right) = \left(u^3,3u^2v,3uv^2,v^3\right) \label{p3ibi} \ , \\
&\left(p_{4a}.p_{4b},p_{4c},p_{4d},p_{4e},p_{4f}\right) =  \notag \\
&\left(u^4,4u^3v,4u^2v^2,2u^2v^2,4uve^3,v^4\right)  \ ,
\label{p4ibi}
\end{align}
where $u\equiv1/\left(e^{\lambda_{11}}+1\right)^2$ and $v=1-u$.
Substituting these into (\ref{11eq5}) and (\ref{11eq6}) and noting that $3Q_3+4Q_4=\kbar$, the edge OPs are also found to be binomial
\beq
e_{11} = u^2 \quad {\rm and} \quad e_{DD} = v^2 \Rightarrow e_{1D} = 2uv \ . 
\label{ebi}
\eeq
From eqs. (\ref{11eq7}), (\ref{Le2}), and (\ref{ebi}), we also have
\beq
e_{11}+\frac{1}{2}e_{1D} = \frac{1}{1 + e^{\nu_2/2}} = \frac{p\bar{z}_1}{\bar{z}} = u \ ,
\label{ejmbi}
\eeq 
and consequently $e^{-\lambda_{11}}=u/v = e^{-\nu_2/2}$. 
Results (\ref{p3ibi})-(\ref{ebi}) are remarkable - the OPs $p_{3i}$, $p_{4i}$, and $e_{jm}$ are all binomial in $u$ and $v$. This substantiates and extends a conjecture in~\cite{Bl25}. Moreover, relation (\ref{ejmbi}) demonstrates that $u$ and $v$ have a direct physical meaning - they are, respectively, the fraction of the contacts per disc of the small and large discs.

Finally, applying the above to the disorder criterion (\ref{criteria}) yields the desired fraction of $3$-cells in the ideal COD,$Q_{3,max}$,
\beq
Q_{3,max} = {\rm min}\left\{\frac{A}{u^3},\frac{A}{(1-u)^3},1\right\}  \ .
\label{Q3max1}
\eeq
This criterion leads to an exponentially decaying probability of the ordered clusters with their size, as shown in the supplemental material~\cite{SM}, with the probability that such a cluster is larger than six identical $3a$-cells lower than $10^{-3}$ for any combination of $p$ and $D$. 
As evident from the above relations, $Q_3$, which depends only on $u$, determines directly $Q_4$, $\kbar$, and $\bar{z}$: $Q_4=1-Q_3$, $\kbar=4-Q_3$, and $\bar{z}=2\kbar/(\kbar-2)$. Thus, to calculate them explicitly, the next task is to relate $u$ to $p$ and $D$. \\

\noindent\underline{\textbf{The dependence of $u$ on $p$ and $D$} }

To relate $Q_3$ to the given parameters $p$ and $D$, using (\ref{Q3max}), requires deriving the function $u(p,D)$. To this end, consider the statistics of the partial areas that the discs occupy within each cell. While the disc area in each $3$-cell configuration is fixed, $S_{3i}^{disc}$, the mean area of a $4$-cell configuration is an average over one of the internal configuration's angles,  
\beq
\langle S_{4i}^{disc}\rangle = \int\limits_{\alpha_{min,4i}}^{\alpha_{max,4i}} S_{4i}^{disc}(\alpha) \ d\alpha \ ,
\eeq
with $\alpha_{min,4i}$ and $\alpha_{max,4i}$ listed in Table~\ref{table1}, the total disc area is then 
\begin{align} 
\bar{S}^{disc} = N_c\left[Q_3\sum_{i=a}^d p_{3i} S_{3i}^{disc} + Q_4\sum_{i=a}^f p_{4i} \langle S_{4i}^{disc}\rangle \right] \ .
\label{sdisc1}
\end{align} 

The $3$-cell configurations disc areas are:
\begin{align}
S^{disc}_{3a} & = \frac{\pi}{8} \quad ; \quad S^{disc}_{3b} = \frac{AC+D^2 AS}{4}  \notag \\
S^{disc}_{3c} & =  \frac{D^2 BC+BS}{4} \quad ; \quad S^{disc}_{3d} =  \frac{\pi D^2}{8} 
\label{Disc3Areas}
\end{align}
and those of the $4$-cell configurations:
\begin{align}
&S^{disc}_{4a} = \frac{\pi}{4}  \nonumber \\
&S^{disc}_{4b} = \frac{\pi + \beta\left(D^2 - 1\right)}{4}  \nonumber \\
&S^{disc}_{4c} = \frac{\pi D^2}{4} - \frac{\left(D^2-1\right)}{8} (\alpha + \beta)  \nonumber \\
&S^{disc}_{4d} = \frac{1}{4}\left[\pi+\alpha\left(D^2-1\right)\right] \nonumber \\
&S^{disc}_{4e} = \frac{1}{8}\left[2\pi D^2 - \beta\left(D^2-1\right)\right]  \nonumber \\
&S^{disc}_{4f} = \frac{\pi D^2}{4} \ ,
\label{eq:4DiscAreas}
\end{align}
with the angles $\beta$ defined in Fig.~\ref{fig:4combs} for each configuration.  \\
All the mean disc areas can be obtained explicitly except for $\langle S^{disc}_{4c}\rangle$:
\begin{align}
\langle S^{disc}_{4a}\rangle &= \frac{\pi}{4} \notag \\
\langle S^{disc}_{4b}\rangle &= \frac{1}{8}\left[2\pi + 3\left(D^2-1\right)AS \right] \notag \\
\langle S^{disc}_{4d}\rangle &= \frac{1}{4}\Bigg\{\pi + \left(D^2-1\right)(AS + BC)\Bigg\} \notag \\
\langle S^{disc}_{4e}\rangle &= \frac{1}{8}\left[2\pi D^2 - 3\left(D^2-1\right)BS \right] \notag \\
\langle S^{disc}_{4f}\rangle &= \frac{\pi D^2}{4} \ .
\label{eq:S4IndDiscAreas}
\end{align}
The mean $\langle S^{disc}_{4c}\rangle$ cannot be expressed in terms of simple functions. Nevertheless, defining $\xi\equiv(1+D)/2$ and substituting the explicit dependence of $\beta$ on $\alpha$,
\begin{align}
\beta(\alpha) = &\arccos\!\left(\frac{1-\xi\cos\alpha}{\sqrt{1+\xi^2-2s\cos\alpha}}\right) + \notag \\
&\arccos\!\left(\frac{2-\xi-\cos\alpha}{\sqrt{1+\xi^2-2\xi\cos\alpha}}\right) \ ,
\end{align}
makes it possible to compute it numerically. This calculation has been carried out for $D=1,1.5,2,2.5,3,3.5,4,4.5,5,5.5,6$, yielding $\langle S_{4c}^{disc}\rangle=0.7854$, $1.2103$, $1.6964$, $2.2387$, $2.8326$, $3.4743$, $4.1606$, $4.8888$, $5.6568$, $6.4625$, $7.3042$, respectively. 

\begin{figure}[h]
\includegraphics[width=1.\linewidth]{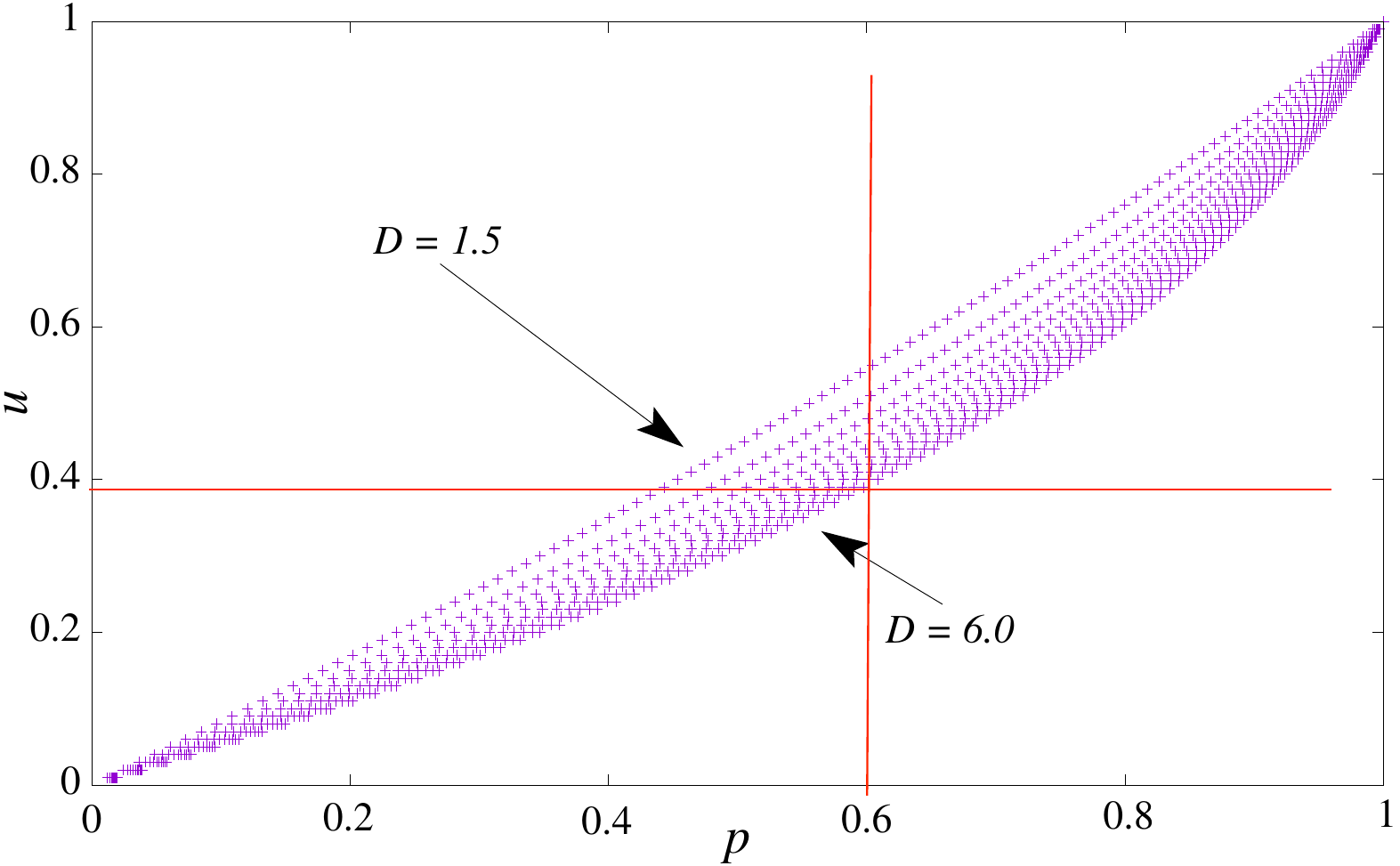}
\caption{The dependence of $u$ on $p$ given $D$, $u(p|D)$, described by eq.~(\ref{pu}), for several values of $D<D_{max}$. }
\label{fig:pu}
\end{figure}

\begin{figure}[tp]
\includegraphics[width=1.\linewidth]{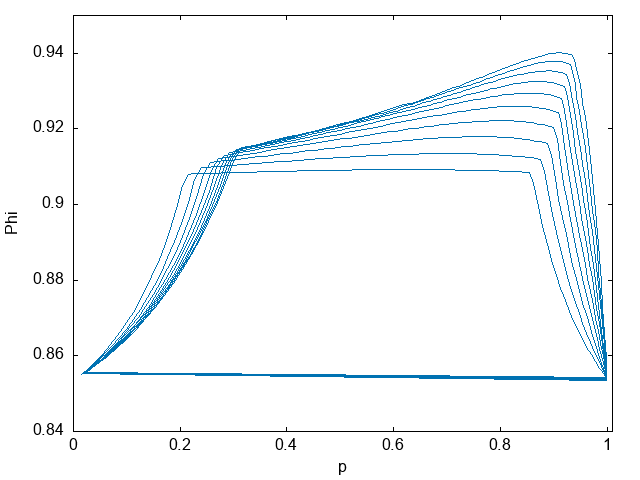}
\caption{$\phi_{max}$ for 11 size ratios $1.0\leq D\leq6.0$.The central region corresponds to $Q_{3,max}=1$.}
\label{phimax}
\end{figure}

As the OPs depend only on $u$ ((\ref{p3ibi})-(\ref{ebi})) and the mean disc areas depend only on $D$, substituting these in expression (\ref{sdisc1}) yields $\bar{S}^{disc}(u,D)$. However, the total disc area is also $N\left[p+(1-p)D^2\right]\pi/4$. Using relation (\ref{Euler3}) for $N/N_c$ and comparing to (\ref{sdisc1}), provides the desired relation $p(u,D)$ for the densest packing,
\beq
p = \frac{1}{D^2-1}\left[D^2 - \frac{8 \bar{S}^{disc}(u,D)}{(\kbar - 2)\pi}\right] \ .
\label{pu}
\eeq
For $D=1$, $u=p$ exactly and for the 10 values of $1<D<D_{max}$ calculated above, the functions $u(p|D)$ are shown in Fig.~\ref{fig:pu}. 
Specialising relation (\ref{Genphi}) for $\phi_{max}$,
\beq
\phi_{max} = \frac{Q_{3,max}\bar{S}_3^{disc} + \left(1-Q_{3,max}\right)\bar{S}_4^{disc} }{Q_{3,max}\bar{S}_3 + \left(1-Q_{3,max}\right)\bar{S}_4} \ ,
\label{34phi}
\eeq
and using relation (\ref{pu}) to convert from $u$ to $p$, yields the main result of this paper, $\phi_{max}(p,D)$. This function is shown in Fig.~\ref{phimax} for the values of $D$ analysed above, including for $D=1$ as a baseline.
The plots exhibit a central region in which $Q_{3max}=1$, which means that any random packing realisation would be disordered. Interestingly, for any value of $D$ in the range studied here, the highest possible packing fraction is achieved for a specific value of $p$ that lies within this region. A table summarising the results numerically for $1<D<D_{max}$ in steps of $0.5$ and $0.0<p<1.0$ in steps of $0.1$ are shown in table X in the supplemental material~\cite{SM}. \\
 
\begin{figure}[tp]
\includegraphics[width=1.\linewidth]{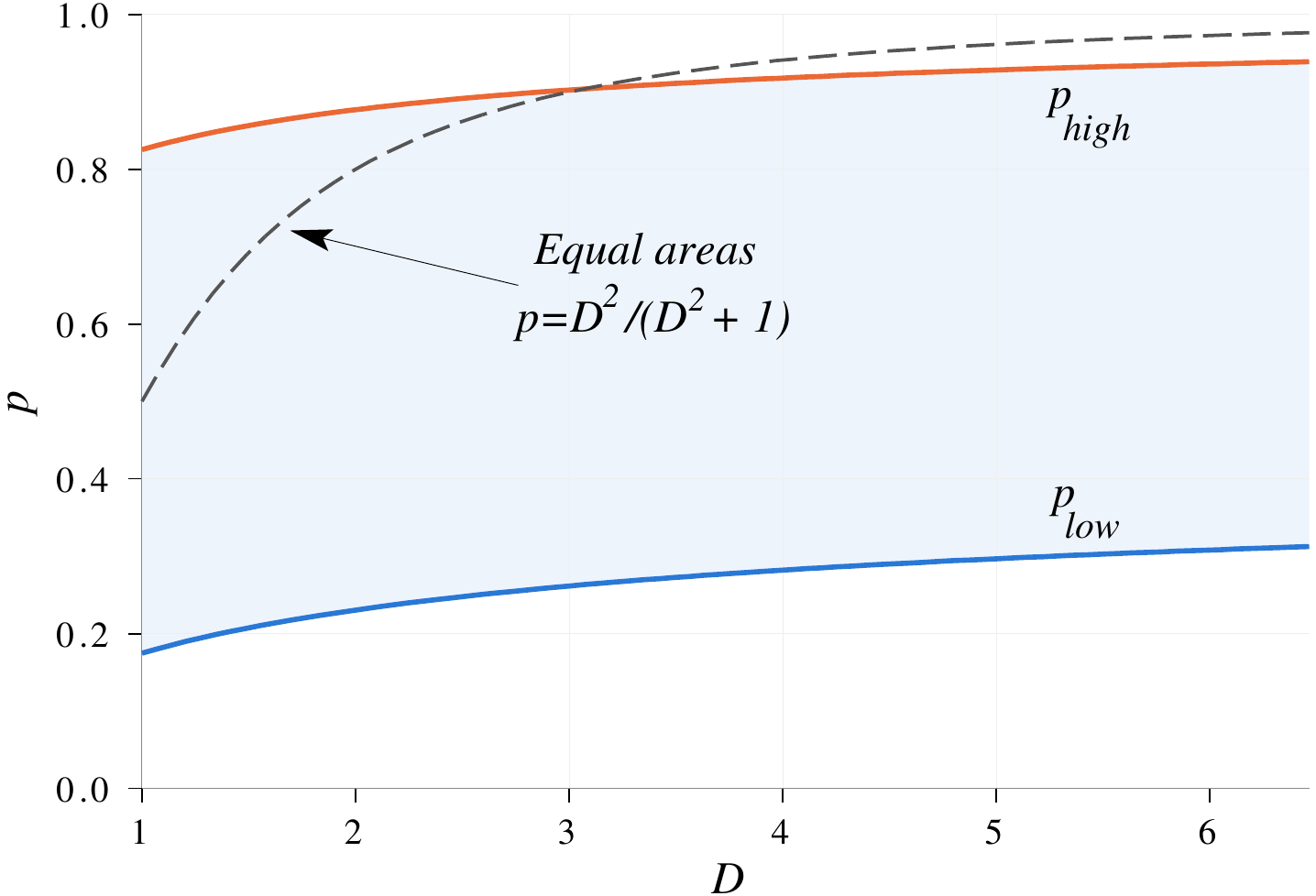}
\caption{Region of assured disorder. For any $D$, $Q_{3,max}=1$ in the shaded region between $p_{low}$ and $p_{high}$ and disorder is assured. 
The dashed line is the value of $p_{high}$ for the commonly used practice of both disc species occupying the same area, in which case $p$ is a function of $D$. 
The plot shows that this practice runs a higher risk of order for $D\lesssim3$, below which $p_{high}$ drops considerably.}
\label{pD}
\end{figure}

Using (\ref{pu}) and (\ref{Q3max}), it is also possible to identify, for any value of $D$, the lowest and highest values of $p$, between which bidisperse packings are bound to be disordered. These are shown in Fig.~\ref{pD}. Also shown in the figure is the commonly used special case of both disc species occupying the same disc area, which couples $p$ and $D$, $p=D^2/\left(D^2+1\right)$. In this case, the highest safe value of $p$ drops rapidly below $D\lesssim3$, an observation that should inform designs of experiments and simulations to avoid order.
For any value of $D$, $Q_{3,max}=1$ in the region between $p_{low}$ and $p_{high}$, i.e., the packing is disordered even if it contains only $3$-cells. 
$p_{low}$ runs from $0.1746$ at $D=1$ to $0.3123$ at $D=D_{max}$ and $p_{high}$ runs from $0.8254$ at $D=1$ to  $0.9390$ at $D=D_{max}$.
The bounded region between $p_{low}$ and $p_{high}$ corresponds to the central range of $p$ in Fig.~\ref{phimax}. Within this region, the value of $w$ in (\ref{criteria}) is imaterial and it does not affect either the OPs $p_{3i}$ ($i=a,b,c,d$) or $\phi_{max}$. A discussion on the best way to choose $w$ outside this range, is presented in the concluding section. Also shown in Fig.~\ref{pD} is the highest disorder-ensuring value of $p$, when $w=1/3$, for the common choice of bidisperse packings with the two disc species occupying the same total area. In these packings, $p$ and $D$ are related and $p_{high}$ dips significantly for $D\lesssim3$, reducing the range in which disorder is certain.
Plots~\ref{phimax} and \ref{pD} should be useful in planning compositions of very dense bidisperse packings in numerical and physical experiments. \\

\noindent\underline{\textbf{Summary and conclusions} }

To conclude, a theory, based on the cell order distribution (COD), has been developed to predict the highest possible packing fraction of bidisperse disc packings in two dimensions $\phi_{max}$. The discs size ratio was limited to $D_{max}=\sqrt{3}/(2-\sqrt{3})$ to exclude rattlers in $3$-cells and the number fraction of the two disc species was parametererised by that of the small discs, $p$. 
Having shown that the lower the mean cell order, $\kbar$, the higher the packing fraction, the analysis focused on disordered packings of only $3$- and $4$-cells as the ones that could yield the highest $\phi_{max}$. 
To ensure that the studied packings are disordered, a disorder criterion has been presented, which limits the probability of adjacent equilateral cells. This has been shown to force the occurrence probabilities (OPs) of ordered clusters to decay exponentially with size, $L$. One of the advantages of this criterion is that it makes it possible to determine, for any disc size ratio, the range of $p$, within which the packing is assured to be disordered. This a-priori criterion is independent of the packing protocol, obviating the difficulty of the difficult-to-explore infinite parameter space of all possible packing protocols. 
It has been shown in the supplemental material~\cite{SM} that the adjacency probability of equilateral triangles, $w$, only affects the rate of exponential decay of ordered clusters probability within relatively small ranges of $p$ close to $0$ and $1$. In a wide region, $p_{low}\leq p\leq p_{high}$, $w$ has no effect whatever. For the smaller regions $0<p<p_{low}$ and $p_{high}<p<1$ $w$ should be chosen such that the OP of a cluster of a chosen size, say $L=6$, is below a prescribed threshold. This has been demonstrated for $p=0.5$ - the OP of clusters of equilateral cells of $L=6$ is below $10^{-3}$ and neither the cell configuration probabilities nor $\phi_{max}$ depend on $w$.  
 
The OPs of all the configurations of $3$- and $4$-cells have been derived for any combination of $p$ and $D$ and were shown to be binomial in a parameter $u$. This parameter also describes the OPs of the three types of inter-disc contacts: $e_{11}$, $e_{1D}$, and $e_{DD}$ and it has been shown to have a physical interpretation - it is the fraction of contacts of the small discs per disc, $u=p\bar{z}_1/\bar{z}$. 
Next, the statistics of cell areas and of the areas, which the discs occupy inside them, were derived and used to express $u$ in terms of $p$ and $D$, as well as to derive $\phi_{max}(p,D)$, achieving the main objective of this paper. The obviation of the protocols infinite parameter space means that he obtained results and, in particular, $\phi_{max}$, are general irrespective of the packing protocol and packing history. 
The value of $\phi_{max}$ has been derived by averaging over the OPs of all the $3$-cell configurations for any $p$ and $D$ and, therefore, it straightforwardly satisfies the Florian upper bound~\cite{Fl60} for any $D$. This is because the Florian bound is simply the specific packing fraction of configuration $3b$ in Fig.~\ref{fig:3combs}, which is the highest of all those configurations. It also satisfies the de Laaf et al. bound in two dimensions~\cite{Laetal14}, which is less less tight than Florian's.

The identification of the range of $p(D)$, within which any packings are assured to be disordered, should assist in designing experiments and simulations that aim to study disordered packings. In passing, it was shown that the common choice of packings of bidisperse discs that occupy the same area increases the risk of order for $D\lesssim3$. 

It should be emphasised that $\phi_{max}(p,D)$ is the highest possible packing fraction that any disordered bidisperse packing can achieve and the derivation here is not concerned with finding the specific protocol that can achieve or even if there is one. Nevertheless, $\phi_{max}(p,D)$ can serve as a guide for future physical and numerical experiments when high packing density is desired. Moreover, measuring packing fractions that exceed this value would be an indication of extended crystalline domains and  eliminate the need to search the structure in detail to discover whether such order exists or not. In this author's view, $\phi_{max}$ could be regarded as the random close packing fraction, $\phi_{RCP}$, whether it is practically achievable or not, but this is likely a matter of the definition of $\phi_{RCP}$.

The method developed here is useful in its own right as eq. (\ref{Genphi}) is general and describes both ordered and disordered structures, given their CODs. For example, it yields the correct packing fraction of each of the ordered bidisperse structures discussed in~\cite{Fe25}, which comprise only $3$-cells and have specific combinations of $p$ and $D$. 

Another potential extension of this approach is to analysis of tridisperse packings of disc sizes, $1$, $D_1$, and $D_2>D_1$, at respective concentrations $p$, $p_1$, and $p_2$. In such systems, the disorder criterion presented here should be applied to the three possible same-size $3$-cell configurations that form equilateral cells. It is intriguing to test whether or not the OPs of those configurations turn out to be trinomial as, if so, the analysis, although more cumbersome, would be conceptually similar.   

This approach provides both a density prediction and predictions for the structure supporting it. It improves on existing theories of bidisperse packing in relating explicitly the maximum packing fraction and the underlying contact and cell statistics. The constrained maximum-entropy calculations yield cell and edge probabilities, governed by the same binomial variable $u$, which has a direct structural interpretation - it is the fraction of the mean coordination of the small discs, $u=p\bar{z}_1/\bar{z}$. This improves on previous estimates of these specific mean coordination numbers, $\bar{z}_1=u\bar{z}/p$ and $\bar{z}_D=(1-u)\bar{z}/(1-p)$. Significantly, this circumvents previous estimates of these quantities via imposing isostatic coordination or fitting a reference packing fraction.
$u(p,D)$ is determined by the total disc area, while the explicit disorder criterion determines the maximal fraction of $3$-cells. Combined with the cell statistics, these yield $\phi_{\max}(p,D)$ within the stated $3$- and $4$-cell description.

\bibliography{}

\clearpage

\onecolumngrid
\setcounter{equation}{0}
\setcounter{table}{0}

\section{Supplemental Material }
 
\underline{\textbf{Occurrence probabilities of crystalline $3$-cell clusters}} \\

The detailed calculation in the main text of the occurrence probabilities (OPs) of $3$-cell configurations $a$ and $d$ are $p_{3a}=u^3$ and $p_{3d}=(1-u)^3$. 
\begin{align}
w \geq &Q_3p_{3a}\left[ \left(Q_3p_{3a}\right)^3 + 3\left(1 - Q_3p_{3a}\right)\left(Q_3p_{3a}\right)^2\right] \notag \\
w \geq &Q_3p_{3d}\left[\left(Q_3p_{3d}\right)^3 + 3\left(1 - Q_3p_{3d}\right)\left[Q_3p_{3d}\right)^2\right] \ .
\label{criteria}
\end{align}
The disorder criterion in the main text  limits the OP of more than one neighbour equilateral triangle to below $w=1/3$, for which $A=0.562236...$.. Here, I first formulate the probability that a randomly chosen equilateral cell is in a cluster of $L$ such cells, $P(L)$. This probability describes the OPs of clusters as a function of their sizes.  I then check the effect of varying $w$ on $P(L)$. The largest fraction, $Q_3$ is determined by eqs. (\ref{criteria}),
\begin{equation}
Q_{3,max} = {\rm min} \left\{ \frac{A}{u^3}, \frac{A}{(1-u)^3}, 1\right] \  .
\label{Q3max}
\end{equation}
The symmetry of the expressions in the two configurations makes it sufficient to analyse only one of them, say, $3a$. 
To determine $P(L)$, start from Euler's relation in the lane:
\begin{equation}
N_v-N_e+N_c=1 \ ,
\label{Euler0}
\end{equation}
and apply it to the cluster. Let the cluster include $N_{i}$ internal edges, which are common to two cells within the cluster, and $N_{eb}=N_{e}-N_{i}$ edges located on the boundaries of the boundary cells. Similarly, $N_v$ include $m$ internal nodes and $N_{vb}=N_{v}-m$ boundary nodes.
The internal edges, which are shared between two cells, connect either two internal nodes or one internal and one boundary node. The  boundary edges are shared only between boundary nodes, unless the two boundaries are only one cell apart. The latter case can happen either when a cluster is small, e.g., in $L=2$, or when it is very elongated. The following analysis focuses on clusters with no holes, which is the case at least up to clusters of $L=12$, as verified by explicit counting, detailed to below in Table~\ref{tableB}. In these clusters $N_{vb}=N_{eb}$., which modifies the above Euler relation,
\begin{equation}
L- N_{i} + m = 1 \ ,
\label{Euler1}
\end{equation}
As every cell has exactly three edges and each edge connects two nodes,
\begin{equation}
3L = 2N_i + N_{vb} \ .
\label{edges}
\end{equation}
It follows that:
\begin{equation}
N_i = L -1 + m \quad ;\quad N_{vb}.= L + 2 - 2m \quad ; \quad N_v = N_{vb} + m =  L + 2 - m \ .
\label{Lmv}
\end{equation}
Note that there can be no internal nodes when $L<6$. 

`Denoting by $X\equiv Q_3p_{3a}$ the OP of a $3a$ configurations, consider starting from a nuclear such a cell and adding adjacent identical cells, one at a time, up to a cluster of $L$ cells. For example, a cluster of $L=2$ is constructed by adding one $3a$ cell to one of the nuclear cell's edges. The resulting $L=2$ cluster has four boundary edges, no internal node and its OP is $P(2)=3X^2(1-X)^4$. Generally, given $L$ and $m$, 
\begin{equation}
P(L) = X^L B_{L,m}(1-X) \ ,
\label{GenOP}
\end{equation}
where $B_{L,m}(y)$ are polynomials that count the different cluster shapes and the term $y\equiv(1-X)$ collects clusters with the same number of non-$3a$ cells on the outside of their boundary edges, $N_p$. Table~\ref{tableB} lists the OPs up to $L=12$.  These values were computed using ChatGPT, using the author's analytical expressions.

\clearpage

{\small
\begin{longtable}{rrrrl}
\caption{Cluster-shape polynomials $B_{L,m}(y)$ for $L\le12$; $y=1-X$. $N_b=L+2-2m$ and $A_{L,m}=B_{L,m}(1)$.}\label{tableB}\label{tab:B}\\
\toprule
$L$ & $m$ & $N_b$ & $A_{L,m}$ & $B_{L,m}(y)$\\\midrule\endfirsthead
\toprule
$L$ & $m$ & $N_b$ & $A_{L,m}$ & $B_{L,m}(y)$\\\midrule\endhead
1 & 0 & 3 & 1 & $y^{3}$\\
2 & 0 & 4 & 3 & $3\,y^{4}$\\
3 & 0 & 5 & 9 & $9\,y^{5}$\\
4 & 0 & 6 & 28 & $12\,y^{5} + 16\,y^{6}$\\
5 & 0 & 7 & 75 & $60\,y^{6} + 15\,y^{7}$\\
6 & 0 & 8 & 189 & $81\,y^{6} + 90\,y^{7} + 18\,y^{8}$\\
6 & 1 & 6 & 3 & $3\,y^{6}$\\
7 & 0 & 9 & 434 & $56\,y^{6} + 231\,y^{7} + 126\,y^{8} + 21\,y^{9}$\\
7 & 1 & 7 & 21 & $21\,y^{7}$\\
8 & 0 & 10 & 984 & $48\,y^{6} + 264\,y^{7} + 480\,y^{8} + 168\,y^{9} + 24\,y^{10}$\\
8 & 1 & 8 & 108 & $72\,y^{7} + 36\,y^{8}$\\
9 & 0 & 11 & 2214 & $216\,y^{7} + 972\,y^{8} + 783\,y^{9} + 216\,y^{10} + 27\,y^{11}$\\
9 & 1 & 9 & 387 & $108\,y^{7} + 270\,y^{8} + 9\,y^{9}$\\
10 & 0 & 12 & 5030 & $120\,y^{7} + 1230\,y^{8} + 2210\,y^{9} + 1170\,y^{10} + 270\,y^{11} + 30\,y^{12}$\\
10 & 1 & 10 & 1155 & $180\,y^{7} + 675\,y^{8} + 300\,y^{9}$\\
10 & 2 & 8 & 15 & $15\,y^{8}$\\
11 & 0 & 13 & 11319 & $1122\,y^{8} + 4092\,y^{9} + 4092\,y^{10} + 1650\,y^{11} + 330\,y^{12} + 33\,y^{13}$\\
11 & 1 & 11 & 3003 & $132\,y^{7} + 1254\,y^{8} + 1287\,y^{9} + 330\,y^{10}$\\
11 & 2 & 9 & 132 & $66\,y^{8} + 66\,y^{9}$\\
12 & 0 & 14 & 25380 & $792\,y^{8} + 5112\,y^{9} + 10044\,y^{10} + 6768\,y^{11} + 2232\,y^{12} + 396\,y^{13} + 36\,y^{14}$\\
12 & 1 & 12 & 7536 & $6\,y^{6} + 108\,y^{7} + 1602\,y^{8} + 3156\,y^{9} + 2304\,y^{10} + 360\,y^{11}$\\
12 & 2 & 10 & 702 & $234\,y^{8} + 396\,y^{9} + 72\,y^{10}$\\
\bottomrule
\end{longtable}}
For example, when $D=3$ and $p=0.5$, for which $Q_{3,max}=1$, $X_a=u^3=0.0486$, and $X_d=(1-u)^3=0.2560$, 
the OPs of $3a$ and $3d$ clusters are shown in Table~\ref{table305} for $w=1/3$.
\begin{table}[htbp]
\centering\small
\caption{Occurrence probabilities of 3a and 3d clusters for $D=3$, $p=0.5$, and $w=1/3$.}\label{table305}
\begin{tabular}{lll}
\toprule
$L$ & $P_L$, 3a clusters & $P_L$, 3d clusters\\\midrule
1 & $4.19\times10^{-2}$ & $1.05\times10^{-1}$\\
2 & $5.81\times10^{-3}$ & $6.02\times10^{-2}$\\
3 & $8.07\times10^{-4}$ & $3.44\times10^{-2}$\\
4 & $1.19\times10^{-4}$ & $2.34\times10^{-2}$\\
5 & $1.50\times10^{-5}$ & $1.33\times10^{-2}$\\
6 & $2.40\times10^{-6}$ & $8.10\times10^{-3}$\\
7 & $3.91\times10^{-7}$ & $4.49\times10^{-3}$\\
8 & $6.91\times10^{-8}$ & $2.73\times10^{-3}$\\
9 & $1.04\times10^{-8}$ & $1.61\times10^{-3}$\\
10 & $1.73\times10^{-9}$ & $9.64\times10^{-4}$\\
11 & $3.02\times10^{-10}$ & $5.53\times10^{-4}$\\
12 & $5.35\times10^{-11}$ & $3.27\times10^{-4}$\\
\midrule
$\sum_{L\le12}P_L/X$ & 1.0000 & 0.998\\
\bottomrule
\end{tabular}
\end{table}
Moreover, practically all the $3a$ cells and $99.8\%$ of the $3d$ cells are in clusters of at most $L=12$.
The enumeration of clusters in Table~\ref{tableB} were produced by polyiamond clusters.py, run as python3 polyiamond clusters.py 12, the output of which is available from the author on request.  \\
\begin{figure}[tp]
\includegraphics[width=.9\linewidth]{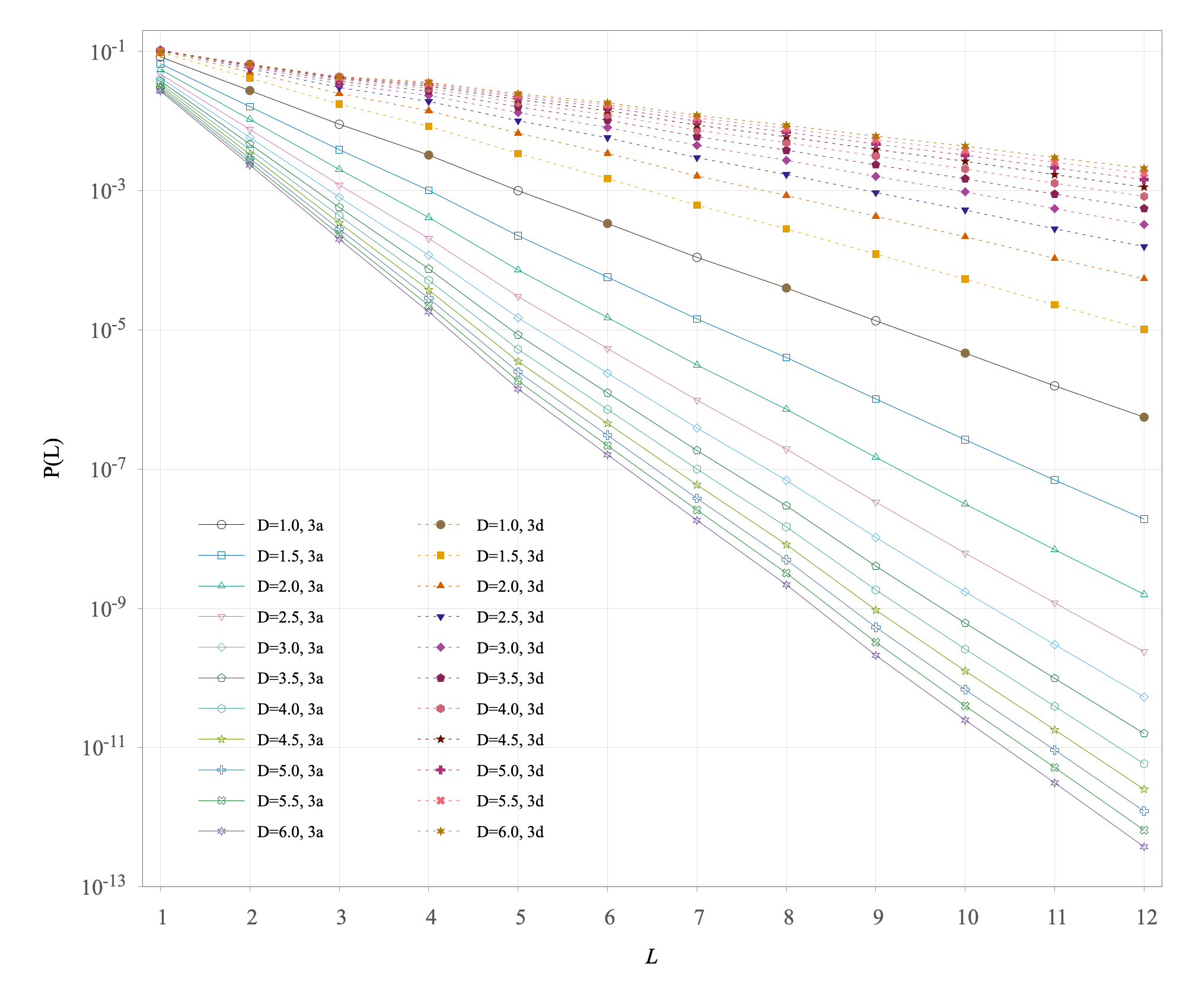}
\caption{$P(L)$ decays almost exactly exponentially with $L$ for $D=1.0, 1.5, 2.0, ..., 6.0$ and $p=0.5$.
Solid and dashed lines show, respectively, the OPs of the $3a$ and $3d$ clusters.}
\label{Pexpo}
\end{figure}
The cluster OPs decay exactly exponentially for both the $3a$ and $3d$ clusters. An example of the decrease of $P(L)$ with $L$ for $D=1.0, 1.5, 2.0, ..., 6.0$ and $p=0.5$ is shown in Fig.~\ref{Pexpo}.   \\

Examining Figs.~4 and 5 in the main text, one can see that, within a wide range of values of $p$, the solution to (\ref{Q3max}) is $Q_{3,max}=1$. In this range, the value of $w$ is immaterial - $P(L)$, the OPs $p_{3i}$ ($i=a,b,c,d$), and $\phi_{max}$ are all independent of its choice. \\

\underline{\textbf{The factor $P$ is positive definite}} \\

This section substantiates the result in the main text, $e^{-\lambda_{11}-\lambda_{DD}}=1$. 
First, express $e_{jm}$ in the text in terms of $e^{-\lambda_{11}}$ and $e^{-\lambda_{DD}}$, using relations (19)-(21) in the main text. 
Then, rewrite relation (39) in the main text as 
\begin{equation}
\left(\kb Z_3 Z_4\right)^2\left[e_{1D}^2 - 4e_{11}e_{DD}\right] = 0 \ ,
\label{A1}
\end{equation}
where $\kb$, $Z_3$, and $Z_4$ were defined there, and plug the expressions for the $e_{jm}$ into it. 
Next, collect numerical coefficients to obtain
\begin{align}
&\left(e_{1D}^2-4e_{11}e_{DD}\right)\left(\kb Z_3 Z_4\right)^2 = \notag \\
&36 Q_3^2 Z_4^2\left[\left(e^{-\lambda_{11}}+e^{-\lambda_{DD}}\right)^2 - \left(e^{-3\lambda_{11}}+e^{-\lambda_{11}}\right)\left(e^{-3\lambda_{DD}}+e^{-\lambda_{DD}}\right)\right] + \notag \\
&48 Q_3 Q_4 Z_3 Z_4 \Bigg\{2\left(e^{-\lambda_{11}}+e^{-\lambda_{DD}}\right)\left(e^{-2\lambda_{11}}+e^{-\lambda_{11}-\lambda_{DD}}+1+e^{-2\lambda_{DD}}\right) \notag - \\
&\left(e^{-3\lambda_{11}}+e^{-\lambda_{11}}\right)\left(e^{-\lambda_{11}-\lambda_{DD}}+2e^{-2\lambda_{DD}}+e^{-4\lambda_{DD}}\right) - \notag \\
&\left(e^{-4\lambda_{11}}+2e^{-2\lambda_{11}}+e^{-\lambda_{11}-\lambda_{DD}}\right)\left(e^{-\lambda_{DD}}+e^{-3\lambda_{DD}}\right)\Bigg\} + \notag \\
&64Q_4^2 Z_3^2\Bigg\{\left(e^{-2\lambda_{11}}+e^{-\lambda_{11}-\lambda_{DD}}+1+e^{-2\lambda_{DD}}\right)^2 - \notag \\
&\left(e^{-4\lambda_{11}}+2e^{-2\lambda_{11}}+e^{-\lambda_{11}-\lambda_{DD}}\right)\left(e^{-\lambda_{11}-\lambda_{DD}}+2e^{-2\lambda_{DD}}+e^{-4\lambda_{DD}}\right)\Bigg\} \ .
\label{A2}
\end{align}
Rearrange the terms in the parentheses. This yields
\begin{align}
\left(e_{1D}^2-4e_{11}e_{DD}\right)\left(\kb Z_3 Z_4\right)^2 = \left(1 - e^{-\lambda_{11}-\lambda_{DD}}\right) P \ ,
\label{A3}
\end{align}
where
\begin{align}
P \equiv 36 Q_3^2 Z_4^2 C_1 + 48 Q_3 Q_4 Z_3 Z_4 C_2 + 64Q_4^2 Z_3^2 C_3  
\label{A4}
\end{align}
and 
\begin{align}
C_1 = &\ e^{-2\left(\lambda_{11}+\lambda_{DD}\right)}+e^{-2\lambda_{11}}+e^{-\lambda_{11}-\lambda_{DD}}+e^{-2\lambda_{DD}} \notag \\
C_2 = & \left(e^{-\lambda_{11}}+e^{-\lambda_{DD}}\right)\left(e^{-2\left(\lambda_{11}+\lambda_{DD}\right)}+2e^{-2\lambda_{11}}+e^{-\lambda_{11}-\lambda_{DD}}+2e^{-2\lambda_{DD}}+2\right) \notag \\
C_3 = &\ e^{-4\lambda_{11}}+e^{-4\lambda_{DD}}+e^{-3\left(\lambda_{11}+\lambda_{DD}\right)}
+2e^{-3\lambda_{11}-\lambda_{DD}}+2e^{-\lambda_{11}-3\lambda_{DD}}+ \notag \\
&\ e^{-2\left(\lambda_{11}+\lambda_{DD}\right)} + 2e^{-2\lambda_{11}}+2e^{-2\lambda_{DD}}+3e^{-\lambda_{11}-\lambda_{DD}}+1 \ .
\label{A5}
\end{align}
$P$ is quadratic in $\left(Q_3,Q_4\right)$ and $C_1$, $C_2$, and $C_3$ are positive definite. Therefore, for any $0<Q_3,Q_4\leq1$, $P$ is positive definite and has no real roots. It follows that 
\begin{equation}
e^{-\lambda_{11}-\lambda_{DD}} - 1 = 0 \ ,
\end{equation}
and, thence,
\begin{equation}
\lambda_{DD} = -\lambda_{11} \ ,
\end{equation}
as stated in the main text. \\

\underline{\textbf{Summary of relevant variables' values for representative combinations of $p$ and $D$} }  \\

\begin{center}\scriptsize\setlength{\tabcolsep}{3pt}
\captionof{table}{The values of $\phi_{max}$, $Q_{3,max}$, $\bar z_1$, $\bar z_D$, and the normalised OPs, $p_{3i}$ and $p_{4i}$ when $Q_3=Q_{3,\max}$, for $D=1.5, 2.0, ..., 6.0$ and $p=0.1, 0.2, ...,0.9$. Note that when $Q_{3,\max}=1$ the $p_{4i}$s are redundant. These data were computed by ChatGPT following the analytical derivations in the main text.}
\label{tableQ}\label{tab:all}
\vspace{4pt}
\begin{tabular}{l|ccccccccc}\toprule
$D=1.5$ & $p=0.1$ & $p=0.2$ & $p=0.3$ & $p=0.4$ & $p=0.5$ & $p=0.6$ & $p=0.7$ & $p=0.8$ & $p=0.9$\\\midrule
$Q_{3,\max}$ & 0.7295 & 0.9776 & 1.0000 & 1.0000 & 1.0000 & 1.0000 & 1.0000 & 1.0000 & 0.8333\\\midrule
$p_{3a}$ & 0.0006 & 0.0048 & 0.0170 & 0.0426 & 0.0882 & 0.1621 & 0.2748 & 0.4396 & 0.6747\\
$p_{3b}$ & 0.0190 & 0.0707 & 0.1472 & 0.2380 & 0.3298 & 0.4056 & 0.4436 & 0.4157 & 0.2837\\
$p_{3c}$ & 0.2097 & 0.3493 & 0.4256 & 0.4437 & 0.4112 & 0.3383 & 0.2388 & 0.1310 & 0.0398\\
$p_{3d}$ & 0.7708 & 0.5751 & 0.4102 & 0.2757 & 0.1709 & 0.0940 & 0.0428 & 0.0138 & 0.0019\\\midrule
$p_{4a}$ & 0.0000 & 0.0008 & 0.0044 & 0.0149 & 0.0392 & 0.0884 & 0.1786 & 0.3342 & 0.5918\\
$p_{4b}$ & 0.0021 & 0.0159 & 0.0505 & 0.1108 & 0.1957 & 0.2949 & 0.3846 & 0.4214 & 0.3317\\
$p_{4c}$ & 0.0232 & 0.0784 & 0.1459 & 0.2066 & 0.2440 & 0.2459 & 0.2070 & 0.1328 & 0.0465\\
$p_{4d}$ & 0.0116 & 0.0392 & 0.0729 & 0.1033 & 0.1220 & 0.1230 & 0.1035 & 0.0664 & 0.0232\\
$p_{4e}$ & 0.2563 & 0.3874 & 0.4217 & 0.3851 & 0.3042 & 0.2051 & 0.1114 & 0.0419 & 0.0065\\
$p_{4f}$ & 0.7067 & 0.4783 & 0.3047 & 0.1794 & 0.0948 & 0.0428 & 0.0150 & 0.0033 & 0.0002\\\midrule
$\phi_{\max}$ & 0.8697 & 0.9040 & 0.9084 & 0.9088 & 0.9091 & 0.9092 & 0.9092 & 0.9088 & 0.8826\\
$\bar z_1$ & 4.280 & 4.977 & 5.140 & 5.237 & 5.341 & 5.452 & 5.572 & 5.703 & 5.290\\
$\bar z_D$ & 5.245 & 6.146 & 6.369 & 6.508 & 6.659 & 6.821 & 6.998 & 7.190$^*$ & 6.672\\
\bottomrule\end{tabular}\\[8pt]
\begin{tabular}{l|ccccccccc}\toprule
$D=2$ & $p=0.1$ & $p=0.2$ & $p=0.3$ & $p=0.4$ & $p=0.5$ & $p=0.6$ & $p=0.7$ & $p=0.8$ & $p=0.9$\\\midrule
$Q_{3,\max}$ & 0.7082 & 0.9184 & 1.0000 & 1.0000 & 1.0000 & 1.0000 & 1.0000 & 1.0000 & 0.8947\\\midrule
$p_{3a}$ & 0.0004 & 0.0034 & 0.0124 & 0.0320 & 0.0684 & 0.1303 & 0.2301 & 0.3862 & 0.6284\\
$p_{3b}$ & 0.0152 & 0.0580 & 0.1237 & 0.2065 & 0.2966 & 0.3802 & 0.4362 & 0.4324 & 0.3158\\
$p_{3c}$ & 0.1905 & 0.3264 & 0.4103 & 0.4437 & 0.4285 & 0.3697 & 0.2756 & 0.1613 & 0.0529\\
$p_{3d}$ & 0.7939 & 0.6122 & 0.4535 & 0.3177 & 0.2064 & 0.1198 & 0.0580 & 0.0201 & 0.0030\\\midrule
$p_{4a}$ & 0.0000 & 0.0005 & 0.0029 & 0.0102 & 0.0280 & 0.0661 & 0.1410 & 0.2813 & 0.5383\\
$p_{4b}$ & 0.0015 & 0.0117 & 0.0382 & 0.0875 & 0.1618 & 0.2570 & 0.3564 & 0.4198 & 0.3606\\
$p_{4c}$ & 0.0188 & 0.0657 & 0.1267 & 0.1879 & 0.2337 & 0.2499 & 0.2252 & 0.1567 & 0.0604\\
$p_{4d}$ & 0.0094 & 0.0328 & 0.0634 & 0.0939 & 0.1169 & 0.1250 & 0.1126 & 0.0783 & 0.0302\\
$p_{4e}$ & 0.2352 & 0.3695 & 0.4203 & 0.4037 & 0.3377 & 0.2430 & 0.1423 & 0.0585 & 0.0101\\
$p_{4f}$ & 0.7351 & 0.5198 & 0.3485 & 0.2168 & 0.1219 & 0.0591 & 0.0225 & 0.0055 & 0.0004\\\midrule
$\phi_{\max}$ & 0.8684 & 0.8958 & 0.9104 & 0.9115 & 0.9124 & 0.9132 & 0.9135 & 0.9131 & 0.8948\\
$\bar z_1$ & 3.774 & 4.299 & 4.634 & 4.764 & 4.909 & 5.070 & 5.253 & 5.462 & 5.348\\
$\bar z_D$ & 5.243 & 6.048 & 6.586 & 6.824 & 7.091 & 7.395 & 7.744 & 8.153 & 8.061\\
\bottomrule\end{tabular}\\[8pt]
\begin{tabular}{l|ccccccccc}\toprule
$D=2.5$ & $p=0.1$ & $p=0.2$ & $p=0.3$ & $p=0.4$ & $p=0.5$ & $p=0.6$ & $p=0.7$ & $p=0.8$ & $p=0.9$\\\midrule
$Q_{3,\max}$ & 0.6952 & 0.8829 & 1.0000 & 1.0000 & 1.0000 & 1.0000 & 1.0000 & 1.0000 & 0.9553\\\midrule
$p_{3a}$ & 0.0003 & 0.0027 & 0.0099 & 0.0260 & 0.0565 & 0.1099 & 0.1993 & 0.3460 & 0.5885\\
$p_{3b}$ & 0.0131 & 0.0504 & 0.1088 & 0.1852 & 0.2723 & 0.3587 & 0.4257 & 0.4406 & 0.3413\\
$p_{3c}$ & 0.1779 & 0.3101 & 0.3974 & 0.4402 & 0.4372 & 0.3900 & 0.3031 & 0.1870 & 0.0660\\
$p_{3d}$ & 0.8087 & 0.6368 & 0.4838 & 0.3486 & 0.2340 & 0.1414 & 0.0719 & 0.0265 & 0.0043\\\midrule
$p_{4a}$ & 0.0000 & 0.0004 & 0.0021 & 0.0077 & 0.0217 & 0.0527 & 0.1164 & 0.2429 & 0.4932\\
$p_{4b}$ & 0.0012 & 0.0094 & 0.0312 & 0.0732 & 0.1393 & 0.2291 & 0.3316 & 0.4124 & 0.3813\\
$p_{4c}$ & 0.0162 & 0.0578 & 0.1139 & 0.1738 & 0.2237 & 0.2491 & 0.2360 & 0.1750 & 0.0737\\
$p_{4d}$ & 0.0081 & 0.0289 & 0.0570 & 0.0869 & 0.1119 & 0.1246 & 0.1180 & 0.0875 & 0.0369\\
$p_{4e}$ & 0.2210 & 0.3558 & 0.4160 & 0.4130 & 0.3592 & 0.2709 & 0.1680 & 0.0743 & 0.0142\\
$p_{4f}$ & 0.7534 & 0.5478 & 0.3798 & 0.2454 & 0.1442 & 0.0736 & 0.0299 & 0.0079 & 0.0007\\\midrule
$\phi_{\max}$ & 0.8678 & 0.8918 & 0.9119 & 0.9136 & 0.9152 & 0.9167 & 0.9177 & 0.9178 & 0.9085\\
$\bar z_1$ & 3.462 & 3.898 & 4.300 & 4.443 & 4.605 & 4.791 & 5.007 & 5.265 & 5.428\\
$\bar z_D$ & 5.244 & 6.002 & 6.729 & 7.038 & 7.395 & 7.814 & 8.317 & 8.939 & 9.442\\
\bottomrule\end{tabular}\\[8pt]
\begin{tabular}{l|ccccccccc}\toprule
$D=3$ & $p=0.1$ & $p=0.2$ & $p=0.3$ & $p=0.4$ & $p=0.5$ & $p=0.6$ & $p=0.7$ & $p=0.8$ & $p=0.9$\\\midrule
$Q_{3,\max}$ & 0.6864 & 0.8592 & 1.0000 & 1.0000 & 1.0000 & 1.0000 & 1.0000 & 1.0000 & 1.0000\\\midrule
$p_{3a}$ & 0.0003 & 0.0023 & 0.0084 & 0.0221 & 0.0486 & 0.0959 & 0.1771 & 0.3149 & 0.5542\\
$p_{3b}$ & 0.0116 & 0.0453 & 0.0986 & 0.1700 & 0.2538 & 0.3409 & 0.4148 & 0.4439 & 0.3615\\
$p_{3c}$ & 0.1690 & 0.2981 & 0.3869 & 0.4357 & 0.4415 & 0.4038 & 0.3238 & 0.2085 & 0.0786\\
$p_{3d}$ & 0.8191 & 0.6544 & 0.5061 & 0.3722 & 0.2560 & 0.1594 & 0.0843 & 0.0327 & 0.0057\\\midrule
$p_{4a}$ & 0.0000 & 0.0003 & 0.0017 & 0.0062 & 0.0178 & 0.0439 & 0.0995 & 0.2143 & 0.4553\\
$p_{4b}$ & 0.0010 & 0.0080 & 0.0267 & 0.0636 & 0.1235 & 0.2081 & 0.3106 & 0.4027 & 0.3959\\
$p_{4c}$ & 0.0145 & 0.0524 & 0.1048 & 0.1631 & 0.2149 & 0.2465 & 0.2425 & 0.1892 & 0.0861\\
$p_{4d}$ & 0.0073 & 0.0262 & 0.0524 & 0.0815 & 0.1074 & 0.1232 & 0.1212 & 0.0946 & 0.0430\\
$p_{4e}$ & 0.2109 & 0.3450 & 0.4111 & 0.4179 & 0.3738 & 0.2919 & 0.1893 & 0.0889 & 0.0187\\
$p_{4f}$ & 0.7664 & 0.5681 & 0.4033 & 0.2677 & 0.1626 & 0.0864 & 0.0369 & 0.0104 & 0.0010\\\midrule
$\phi_{\max}$ & 0.8674 & 0.8894 & 0.9129 & 0.9150 & 0.9172 & 0.9194 & 0.9212 & 0.9222 & 0.9202\\
$\bar z_1$ & 3.247 & 3.629 & 4.061 & 4.210 & 4.380 & 4.578 & 4.813 & 5.103 & 5.476\\
$\bar z_D$ & 5.245 & 5.976 & 6.831 & 7.193 & 7.620 & 8.133 & 8.769 & 9.589 & 10.715\\
\bottomrule\end{tabular}\\[8pt]
\begin{tabular}{l|ccccccccc}\toprule
$D=3.5$ & $p=0.1$ & $p=0.2$ & $p=0.3$ & $p=0.4$ & $p=0.5$ & $p=0.6$ & $p=0.7$ & $p=0.8$ & $p=0.9$\\\midrule
$Q_{3,\max}$ & 0.6800 & 0.8420 & 1.0000 & 1.0000 & 1.0000 & 1.0000 & 1.0000 & 1.0000 & 1.0000\\\midrule
$p_{3a}$ & 0.0002 & 0.0020 & 0.0073 & 0.0194 & 0.0431 & 0.0858 & 0.1604 & 0.2904 & 0.5256\\
$p_{3b}$ & 0.0106 & 0.0416 & 0.0911 & 0.1586 & 0.2394 & 0.3262 & 0.4045 & 0.4444 & 0.3771\\
$p_{3c}$ & 0.1623 & 0.2887 & 0.3782 & 0.4312 & 0.4436 & 0.4134 & 0.3399 & 0.2266 & 0.0902\\
$p_{3d}$ & 0.8268 & 0.6677 & 0.5233 & 0.3908 & 0.2739 & 0.1746 & 0.0952 & 0.0385 & 0.0072\\\midrule
$p_{4a}$ & 0.0000 & 0.0003 & 0.0014 & 0.0052 & 0.0151 & 0.0379 & 0.0872 & 0.1923 & 0.4241\\
$p_{4b}$ & 0.0009 & 0.0070 & 0.0236 & 0.0568 & 0.1119 & 0.1919 & 0.2930 & 0.3924 & 0.4057\\
$p_{4c}$ & 0.0133 & 0.0485 & 0.0979 & 0.1546 & 0.2073 & 0.2431 & 0.2463 & 0.2001 & 0.0970\\
$p_{4d}$ & 0.0066 & 0.0242 & 0.0490 & 0.0773 & 0.1037 & 0.1216 & 0.1231 & 0.1001 & 0.0485\\
$p_{4e}$ & 0.2031 & 0.3364 & 0.4064 & 0.4203 & 0.3841 & 0.3081 & 0.2070 & 0.1021 & 0.0232\\
$p_{4f}$ & 0.7761 & 0.5836 & 0.4217 & 0.2857 & 0.1779 & 0.0976 & 0.0435 & 0.0130 & 0.0014\\\midrule
$\phi_{\max}$ & 0.8671 & 0.8878 & 0.9135 & 0.9160 & 0.9187 & 0.9214 & 0.9240 & 0.9258 & 0.9247\\
$\bar z_1$ & 3.089 & 3.435 & 3.883 & 4.033 & 4.207 & 4.411 & 4.657 & 4.967 & 5.380\\
$\bar z_D$ & 5.246 & 5.959 & 6.907 & 7.311 & 7.793 & 8.384 & 9.133 & 10.133 & 11.580\\
\bottomrule\end{tabular}\\[8pt]
\begin{tabular}{l|ccccccccc}\toprule
$D=4$ & $p=0.1$ & $p=0.2$ & $p=0.3$ & $p=0.4$ & $p=0.5$ & $p=0.6$ & $p=0.7$ & $p=0.8$ & $p=0.9$\\\midrule
$Q_{3,\max}$ & 0.6751 & 0.8290 & 1.0000 & 1.0000 & 1.0000 & 1.0000 & 1.0000 & 1.0000 & 1.0000\\\midrule
$p_{3a}$ & 0.0002 & 0.0018 & 0.0066 & 0.0175 & 0.0390 & 0.0782 & 0.1475 & 0.2707 & 0.5009\\
$p_{3b}$ & 0.0099 & 0.0389 & 0.0854 & 0.1497 & 0.2279 & 0.3140 & 0.3950 & 0.4433 & 0.3895\\
$p_{3c}$ & 0.1571 & 0.2812 & 0.3710 & 0.4269 & 0.4443 & 0.4203 & 0.3526 & 0.2420 & 0.1009\\
$p_{3d}$ & 0.8329 & 0.6782 & 0.5371 & 0.4059 & 0.2887 & 0.1876 & 0.1049 & 0.0440 & 0.0087\\\midrule
$p_{4a}$ & 0.0000 & 0.0002 & 0.0012 & 0.0045 & 0.0132 & 0.0334 & 0.0779 & 0.1751 & 0.3978\\
$p_{4b}$ & 0.0008 & 0.0063 & 0.0213 & 0.0518 & 0.1030 & 0.1790 & 0.2783 & 0.3823 & 0.4124\\
$p_{4c}$ & 0.0124 & 0.0455 & 0.0926 & 0.1478 & 0.2009 & 0.2396 & 0.2484 & 0.2087 & 0.1069\\
$p_{4d}$ & 0.0062 & 0.0228 & 0.0463 & 0.0739 & 0.1004 & 0.1198 & 0.1242 & 0.1044 & 0.0534\\
$p_{4e}$ & 0.1970 & 0.3294 & 0.4020 & 0.4215 & 0.3916 & 0.3208 & 0.2217 & 0.1139 & 0.0277\\
$p_{4f}$ & 0.7836 & 0.5958 & 0.4366 & 0.3006 & 0.1908 & 0.1074 & 0.0495 & 0.0156 & 0.0018\\\midrule
$\phi_{\max}$ & 0.8669 & 0.8866 & 0.9139 & 0.9167 & 0.9196 & 0.9228 & 0.9260 & 0.9288 & 0.9287\\
$\bar z_1$ & 2.968 & 3.288 & 3.743 & 3.894 & 4.069 & 4.276 & 4.529 & 4.851 & 5.294\\
$\bar z_D$ & 5.247 & 5.948 & 6.967 & 7.404 & 7.931 & 8.586 & 9.433 & 10.594 & 12.350\\
\bottomrule\end{tabular}\\[8pt]
\begin{tabular}{l|ccccccccc}\toprule
$D=4.5$ & $p=0.1$ & $p=0.2$ & $p=0.3$ & $p=0.4$ & $p=0.5$ & $p=0.6$ & $p=0.7$ & $p=0.8$ & $p=0.9$\\\midrule
$Q_{3,\max}$ & 0.6712 & 0.8188 & 1.0000 & 1.0000 & 1.0000 & 1.0000 & 1.0000 & 1.0000 & 1.0000\\\midrule
$p_{3a}$ & 0.0002 & 0.0016 & 0.0060 & 0.0160 & 0.0358 & 0.0722 & 0.1372 & 0.2545 & 0.4794\\
$p_{3b}$ & 0.0093 & 0.0367 & 0.0809 & 0.1425 & 0.2185 & 0.3036 & 0.3864 & 0.4413 & 0.3994\\
$p_{3c}$ & 0.1528 & 0.2750 & 0.3648 & 0.4230 & 0.4444 & 0.4255 & 0.3628 & 0.2551 & 0.1109\\
$p_{3d}$ & 0.8377 & 0.6867 & 0.5483 & 0.4185 & 0.3013 & 0.1988 & 0.1135 & 0.0492 & 0.0103\\\midrule
$p_{4a}$ & 0.0000 & 0.0002 & 0.0011 & 0.0040 & 0.0118 & 0.0301 & 0.0708 & 0.1612 & 0.3752\\
$p_{4b}$ & 0.0007 & 0.0058 & 0.0196 & 0.0479 & 0.0960 & 0.1686 & 0.2658 & 0.3728 & 0.4168\\
$p_{4c}$ & 0.0117 & 0.0432 & 0.0883 & 0.1421 & 0.1953 & 0.2362 & 0.2495 & 0.2155 & 0.1157\\
$p_{4d}$ & 0.0058 & 0.0216 & 0.0441 & 0.0711 & 0.0977 & 0.1181 & 0.1248 & 0.1078 & 0.0579\\
$p_{4e}$ & 0.1921 & 0.3235 & 0.3981 & 0.4219 & 0.3972 & 0.3311 & 0.2342 & 0.1246 & 0.0321\\
$p_{4f}$ & 0.7897 & 0.6058 & 0.4488 & 0.3130 & 0.2020 & 0.1160 & 0.0550 & 0.0180 & 0.0022\\\midrule
$\phi_{\max}$ & 0.8667 & 0.8857 & 0.9142 & 0.9171 & 0.9203 & 0.9239 & 0.9276 & 0.9312 & 0.9322\\
$\bar z_1$ & 2.872 & 3.172 & 3.630 & 3.780 & 3.955 & 4.164 & 4.421 & 4.753 & 5.218\\
$\bar z_D$ & 5.248 & 5.940 & 7.016 & 7.480 & 8.044 & 8.754 & 9.684 & 10.989 & 13.040\\
\bottomrule\end{tabular}\\[8pt]
\begin{tabular}{l|ccccccccc}\toprule
$D=5$ & $p=0.1$ & $p=0.2$ & $p=0.3$ & $p=0.4$ & $p=0.5$ & $p=0.6$ & $p=0.7$ & $p=0.8$ & $p=0.9$\\\midrule
$Q_{3,\max}$ & 0.6680 & 0.8105 & 1.0000 & 1.0000 & 1.0000 & 1.0000 & 1.0000 & 1.0000 & 1.0000\\\midrule
$p_{3a}$ & 0.0002 & 0.0015 & 0.0055 & 0.0148 & 0.0333 & 0.0674 & 0.1288 & 0.2409 & 0.4607\\
$p_{3b}$ & 0.0088 & 0.0350 & 0.0772 & 0.1367 & 0.2107 & 0.2947 & 0.3788 & 0.4388 & 0.4074\\
$p_{3c}$ & 0.1493 & 0.2698 & 0.3595 & 0.4194 & 0.4440 & 0.4294 & 0.3712 & 0.2664 & 0.1201\\
$p_{3d}$ & 0.8417 & 0.6937 & 0.5577 & 0.4290 & 0.3120 & 0.2085 & 0.1212 & 0.0539 & 0.0118\\\midrule
$p_{4a}$ & 0.0000 & 0.0002 & 0.0010 & 0.0036 & 0.0107 & 0.0274 & 0.0651 & 0.1499 & 0.3558\\
$p_{4b}$ & 0.0007 & 0.0054 & 0.0182 & 0.0448 & 0.0904 & 0.1599 & 0.2551 & 0.3640 & 0.4195\\
$p_{4c}$ & 0.0111 & 0.0413 & 0.0848 & 0.1374 & 0.1905 & 0.2330 & 0.2499 & 0.2210 & 0.1237\\
$p_{4d}$ & 0.0056 & 0.0206 & 0.0424 & 0.0687 & 0.0953 & 0.1165 & 0.1250 & 0.1105 & 0.0618\\
$p_{4e}$ & 0.1880 & 0.3185 & 0.3946 & 0.4218 & 0.4015 & 0.3395 & 0.2449 & 0.1342 & 0.0364\\
$p_{4f}$ & 0.7947 & 0.6141 & 0.4591 & 0.3236 & 0.2116 & 0.1236 & 0.0600 & 0.0204 & 0.0027\\\midrule
$\phi_{\max}$ & 0.8665 & 0.8849 & 0.9144 & 0.9174 & 0.9208 & 0.9246 & 0.9288 & 0.9330 & 0.9353\\
$\bar z_1$ & 2.793 & 3.077 & 3.537 & 3.687 & 3.861 & 4.070 & 4.329 & 4.667 & 5.149\\
$\bar z_D$ & 5.248 & 5.934 & 7.056 & 7.542 & 8.139 & 8.895 & 9.899 & 11.332 & 13.659\\
\bottomrule\end{tabular}\\[8pt]
\begin{tabular}{l|ccccccccc}\toprule
$D=5.5$ & $p=0.1$ & $p=0.2$ & $p=0.3$ & $p=0.4$ & $p=0.5$ & $p=0.6$ & $p=0.7$ & $p=0.8$ & $p=0.9$\\\midrule
$Q_{3,\max}$ & 0.6653 & 0.8036 & 0.9941 & 1.0000 & 1.0000 & 1.0000 & 1.0000 & 1.0000 & 1.0000\\\midrule
$p_{3a}$ & 0.0002 & 0.0014 & 0.0052 & 0.0139 & 0.0313 & 0.0635 & 0.1219 & 0.2295 & 0.4443\\
$p_{3b}$ & 0.0084 & 0.0336 & 0.0743 & 0.1318 & 0.2040 & 0.2871 & 0.3719 & 0.4360 & 0.4139\\
$p_{3c}$ & 0.1463 & 0.2654 & 0.3550 & 0.4162 & 0.4434 & 0.4324 & 0.3781 & 0.2762 & 0.1285\\
$p_{3d}$ & 0.8451 & 0.6996 & 0.5656 & 0.4381 & 0.3213 & 0.2171 & 0.1281 & 0.0583 & 0.0133\\\midrule
$p_{4a}$ & 0.0000 & 0.0002 & 0.0009 & 0.0033 & 0.0099 & 0.0253 & 0.0604 & 0.1405 & 0.3390\\
$p_{4b}$ & 0.0006 & 0.0050 & 0.0171 & 0.0423 & 0.0857 & 0.1527 & 0.2458 & 0.3559 & 0.4211\\
$p_{4c}$ & 0.0106 & 0.0397 & 0.0819 & 0.1335 & 0.1863 & 0.2300 & 0.2500 & 0.2254 & 0.1308\\
$p_{4d}$ & 0.0053 & 0.0199 & 0.0409 & 0.0667 & 0.0932 & 0.1150 & 0.1250 & 0.1127 & 0.0654\\
$p_{4e}$ & 0.1845 & 0.3141 & 0.3914 & 0.4215 & 0.4049 & 0.3465 & 0.2542 & 0.1428 & 0.0406\\
$p_{4f}$ & 0.7989 & 0.6211 & 0.4677 & 0.3328 & 0.2200 & 0.1305 & 0.0646 & 0.0226 & 0.0032\\\midrule
$\phi_{\max}$ & 0.8663 & 0.8842 & 0.9135 & 0.9176 & 0.9211 & 0.9251 & 0.9296 & 0.9345 & 0.9378\\
$\bar z_1$ & 2.727 & 2.999 & 3.447 & 3.607 & 3.781 & 3.990 & 4.250 & 4.592 & 5.087\\
$\bar z_D$ & 5.249 & 5.929 & 7.061 & 7.595 & 8.219 & 9.015 & 10.083 & 11.633 & 14.218\\
\bottomrule\end{tabular}\\[8pt]
\begin{tabular}{l|ccccccccc}\toprule
$D=6$ & $p=0.1$ & $p=0.2$ & $p=0.3$ & $p=0.4$ & $p=0.5$ & $p=0.6$ & $p=0.7$ & $p=0.8$ & $p=0.9$\\\midrule
$Q_{3,\max}$ & 0.6631 & 0.7978 & 0.9826 & 1.0000 & 1.0000 & 1.0000 & 1.0000 & 1.0000 & 1.0000\\\midrule
$p_{3a}$ & 0.0002 & 0.0013 & 0.0049 & 0.0131 & 0.0296 & 0.0603 & 0.1161 & 0.2197 & 0.4297\\
$p_{3b}$ & 0.0081 & 0.0324 & 0.0718 & 0.1276 & 0.1983 & 0.2804 & 0.3656 & 0.4332 & 0.4192\\
$p_{3c}$ & 0.1438 & 0.2616 & 0.3511 & 0.4132 & 0.4427 & 0.4347 & 0.3839 & 0.2847 & 0.1363\\
$p_{3d}$ & 0.8479 & 0.7047 & 0.5722 & 0.4460 & 0.3294 & 0.2247 & 0.1344 & 0.0624 & 0.0148\\\midrule
$p_{4a}$ & 0.0000 & 0.0001 & 0.0008 & 0.0031 & 0.0092 & 0.0236 & 0.0566 & 0.1326 & 0.3242\\
$p_{4b}$ & 0.0006 & 0.0047 & 0.0163 & 0.0402 & 0.0818 & 0.1466 & 0.2378 & 0.3485 & 0.4218\\
$p_{4c}$ & 0.0103 & 0.0384 & 0.0795 & 0.1300 & 0.1826 & 0.2272 & 0.2497 & 0.2291 & 0.1372\\
$p_{4d}$ & 0.0051 & 0.0192 & 0.0397 & 0.0650 & 0.0913 & 0.1136 & 0.1249 & 0.1145 & 0.0686\\
$p_{4e}$ & 0.1815 & 0.3104 & 0.3886 & 0.4210 & 0.4076 & 0.3524 & 0.2622 & 0.1506 & 0.0446\\
$p_{4f}$ & 0.8025 & 0.6272 & 0.4750 & 0.3408 & 0.2275 & 0.1366 & 0.0688 & 0.0247 & 0.0036\\\midrule
$\phi_{\max}$ & 0.8662 & 0.8837 & 0.9117 & 0.9177 & 0.9213 & 0.9255 & 0.9302 & 0.9356 & 0.9400\\
$\bar z_1$ & 2.671 & 2.933 & 3.357 & 3.539 & 3.712 & 3.921 & 4.181 & 4.526 & 5.031\\
$\bar z_D$ & 5.250 & 5.926 & 7.035 & 7.640 & 8.288 & 9.119 & 10.244 & 11.898 & 14.724\\
\bottomrule\end{tabular}
\end{center}

\end{document}